\documentclass[aps,prb,preprint,superscriptaddress,longbibliography]{revtex4}

\usepackage{amsmath}
\usepackage{epsfig,bm}

\begin{document}

\title{Spin Resonances in Iron-Selenide High-$T_c$ Superconductors by Proximity to Hidden Spin Density Wave}

\author{J.P. Rodriguez}

\affiliation{Department of Physics and Astronomy,
California State University, Los Angeles, California 90032}


\begin{abstract}
Recent inelastic neutron scattering studies
by Pan et al., Nature Communications {\bf 8}, 123 (2017),
find evidence for spin resonances
 in an iron-selenide high-$T_c$ superconductor
that persist  at energies above the quasi-particle gap.
The momenta of such spin excitations form a diamond around the checkerboard wavevector, ${\bm Q}_{\rm AF}$,
that is associated with the square lattice of iron atoms
that makes up the system.
It has been suggested that the  ``hollowed-out'' spin-excitation spectrum
is due to hidden N\'eel order.
We study such a hidden spin-density wave (hSDW) state
 that results from nested Fermi surfaces
at the center and at the corner of the unfolded Brillouin zone.
It emerges within mean field theory
 from an extended Hubbard model
over a square lattice of iron atoms that contain the minimal $d_{xz}$ and $d_{yz}$ orbitals.
Opposing N\'eel order exists over the isotropic
 $d+ = d_{xz} + i d_{yz}$ and $d- = d_{xz} - i d_{yz}$ orbitals.
The dynamical spin susceptibility of the hSDW
is computed within the random phase approximation, at perfect nesting.
Unobservable Goldstone modes that disperse acoustically are found at ${\bm Q}_{\rm AF}$.
A threshold is found in the spectrum of  observable spin excitations 
that forms a ``floating ring'' at ${\bm Q}_{\rm AF}$ also.
The ring threshold
moves down in energy toward zero with increasing Hund's Rule coupling,
while it moves up in energy with increasing magnetic frustration.
Comparison with the normal-state features of the spin-excitation spectrum
shown by electron-doped iron selenide is made.
Also, recent predictions 
 of a Lifshitz transition from the nested Fermi surfaces
to Fermi surface pockets at the corner of the folded Brillouin zone will be discussed.
\end{abstract}

\maketitle

\section{Introduction}
Spin resonances inside the energy gap
that opens at the Fermi level
in the spectrum of quasi-particle excitations
of high-temperature superconductors
are commonly observed\cite{keimer_anderson_95,inosov_09}.
In the case of iron-pnictide superconductors,
they are  predicted to exist just below the quasi-particle energy gap,
$2 \Delta_{\rm SC}$, at the nesting wavevector that connects
hole-type Fermi surfaces at the center of the Brillouin zone 
with electron-type Fermi surfaces
at the corner of the folded Brillouin zone\cite{korshunov_eremin_08,maier_scalapino_08}.
Such predictions are based on $S^{+-}$ Bardeen-Cooper-Schrieffer (BCS) groundstates,
where the sign of Cooper pairs alternates between the hole-type and the electron-type Fermi
surfaces\cite{mazin_08,kuroki_08}.
It is believed that low-energy spin fluctuations that arise from the nested Fermi surfaces
are what bind together electrons into Cooper pairs in the $S^{+-}$ state\cite{graser_09}.
The predicted spin resonances inside of the energy gap,
at the ``stripe'' spin-density wave (SDW) wavevectors,
have indeed been observed in iron-pnictide superconductors
by inelastic neutron scattering\cite{inosov_09}.

Spin resonances have also been observed inside the quasi-particle energy gap
of electron-doped iron-selenide high-temperature superconductors,
but at wavenumbers midway between the ``stripe'' SDW ones
and the checkerboard one that describes 
 N\'eel antiferromagnetism\cite{park_11,friemel_12,davies_16,pan_17,ma_17}.
Electron doping buries the hole bands at the center of the Brillouin zone
below the Fermi level,
leaving only the electron-type Fermi surface pockets
 at the corner of the folded Brillouin zone\cite{zhou_13,peng_14,lee_14,zhao_16}.
Spin resonances are therefore observed 
 in electron-doped iron selenide
in  the absence of nested Fermi surfaces,
which is a puzzle.
Additionally,
recent inelastic neutron scattering studies of iron selenide
that is electron-doped by intercalated organic molecules 
find evidence for spin resonances that persist {\it above} the
quasi-particle energy gap, $2\Delta_{\rm SC}$,
at wavenumbers that form a ``diamond''
 around the checkerboard wavevector\cite{pan_17}, $(\pi/a,\pi/a)$.
Such relatively high-energy spin excitations
very likely persist into the normal state
at temperatures above $T_c$.

Recent theoretical work suggests that
the ``rings'' and ``diamonds'' of spin excitations
observed in electron-doped FeSe at the checkerboard wavevector
are due to proximity to a hidden spin-density wave (hSDW) state\cite{jpr_17,jpr_19,jpr_rm_18}.
Here,
the sign of the ordered magnetic moment alternates between
the principal $d+ = (d_{xz}+id_{yz})/\sqrt{2}$ and $d- = (d_{xz}-id_{yz})/\sqrt{2}$
 orbitals of the iron atom,
 as well as between the
``white'' and the ``black'' sites on the checkerboard of iron atoms\cite{jpr_ehr_09,jpr_10}.
It is the most isotropic one among  a family of hSDW states 
that are related by isospin rotations\cite{jpr_20b}.
The stability of the hSDW is driven by perfectly nested Fermi surfaces 
at the center and at the corner of the unfolded Brillouin zone. (See Fig. \ref{FS0}.)
It has recently been shown by the author and a co-worker that fluctuation-exchange
with Goldstone modes associated with such hidden magnetic order results in
a Lifshitz transition to electron/hole Fermi surface pockets at the corner
of the folded Brillouin zone\cite{jpr_rm_18}. (See Fig. \ref{FS1}.)
A rigid shift in energy of this renormalized electronic structure because of electron doping
away from half filling can bury the hole pockets,
leaving the electron pockets that are observed by 
angle-resolved photoemission spectroscopy (ARPES) in electron-doped iron selenide\cite{jpr_20b}.

Below, we shall reveal the nature of spin excitations in the hidden SDW state
within an extended Hubbard model over a square lattice of iron atoms
that includes only the principal $3 d_{xz}$ and $3 d_{yz}$ orbitals
of iron superconductors\cite{jpr_rm_18}.
In particular,
the dynamical spin susceptibility is computed within a 
Nambu-Gorkov-type\cite{nambu_60,gorkov_58,schrieffer_64}
 random phase approximation (RPA) that accounts for perfect nesting of 
the unrenormalized Fermi surfaces mentioned above.
This calculation is then the two-orbital realization
of Schrieffer, Wen and Zhang's ``spin-bag'' calculation of the dynamical spin susceptibility
for the conventional Hubbard model over 
the square lattice\cite{hirsch_85,schrieffer_wen_zhang_89,singh_tesanovic_90,chubokov_frenkel_92}.
As expected, we recover the Goldstone modes that disperse acoustically from the
nesting wavevector, ${\bm Q}_{\rm AF} = (\pi/a,\pi/a)$.
Such modes have an extremely weak spectral weight in the true-spin channel, however.
(See Table \ref{isospin}.)
A ring of spin excitations at ${\bm Q}_{\rm AF}$
begins at energies above the Goldstone modes
in the true spin channel, on the other hand.
They evolve into a diamond shape at ${\bm Q}_{\rm AF}$
as energy increases above the threshold.
We shall argue that the dynamical spin susceptibility within RPA accounts for
spin excitations in the normal state of electron-doped iron selenide.

\section{Nested Fermi Surfaces in Hubbard Model}
The extended Hubbard model for electron-doped iron selenide and the mean field theory
for the hidden SDW state are introduced below.

\subsection{Electron Hopping over Square Lattice of Iron Atoms}
We keep the $3d_{xz}/3d_{yz}$ orbitals of the iron atoms 
in the following description
of a single layer of heavily electron-doped FeSe. In particular,
consider the isotropic basis of orbitals
$d-=(d_{xz}-id_{yz})/{\sqrt 2}$ and $d+=(d_{xz}+id_{yz})/{\sqrt 2}$.
Kinetic dynamics is governed by the hopping Hamiltonian
\begin{equation}
H_{\rm hop} =  
-\sum_{\langle i,j \rangle} (t_1^{\alpha,\beta} c_{i, \alpha,s}^{\dagger} c_{j,\beta,s} + {\rm h.c.}) 
-\sum_{\langle\langle  i,j \rangle\rangle} (t_2^{\alpha,\beta} c_{i, \alpha,s}^{\dagger} c_{j,\beta,s} + {\rm h.c.}),
\label{hop}
\end{equation}
where  the repeated indices $\alpha$ and $\beta$ are summed over the $d-$ and $d+$ orbitals,
where the repeated index $s$ is summed over electron spin $\uparrow$ and $\downarrow$,
and where $\langle i,j\rangle$ and $\langle\langle i,j\rangle\rangle$
represent nearest neighbor (1) and next-nearest neighbor (2) links on the
square lattice of iron atoms.
Above, $c_{i, \alpha,s}$ and $c_{i, \alpha,s}^{\dagger}$
denote annihilation and creation operators for an
electron of spin $s$ in orbital $\alpha$ at site $i$.
The reflection symmetries in a single layer of FeSe imply
that the above intra-orbital and inter-orbital hopping matrix elements
show $s$-wave and $d$-wave symmetry, respectively\cite{raghu_08,Lee_Wen_08,jpr_mana_pds_14}.
Nearest neighbor hopping matrix elements satisfy
\begin{eqnarray}
t_1^{\pm \pm} ({\bf {\hat x}}) &=& t_1^{\parallel} = t_1^{\pm \pm} ({\bf {\hat y}}) \nonumber\\
t_1^{\pm\mp} ({\bf {\hat x}}) &=& t_1^{\perp} = -t_1^{\pm \mp} ({\bf {\hat y}}),
\label{t1}
\end{eqnarray}
with real $t_1^{\parallel}$ and $t_1^{\perp}$,
while next-nearest neighbor hopping matrix elements satisfy
\begin{eqnarray}
t_2^{\pm \pm} ({\bf {\hat x}}+{\bf {\hat y}}) = \; t_2^{\parallel} &=& t_2^{\pm \pm} ({\bf {\hat y}}-{\bf {\hat x}}) \nonumber\\
t_2^{\pm \mp} ({\bf {\hat x}}+{\bf {\hat y}}) = \pm t_2^{\perp} &=& -t_2^{\pm \mp} ({\bf {\hat y}}-{\bf {\hat x}}),
\label{t2}
\end{eqnarray}
with real $t_2^{\parallel}$ and pure-imaginary $t_2^{\perp}$.

The above hopping Hamiltonian is diagonalized\cite{jpr_rm_18} by 
plane waves of $d_{x(\delta)z}$ and $i d_{y(\delta)z}$ orbitals that are rotated
with respect to the principal axis by a phase shift $\delta({\bm k})$:
\begin{eqnarray}
|{\bm k}, d_{x(\delta)z}\rangle\rangle &=&
{\cal N}^{-1/2} \sum_i e^{i{\bm k}\cdot{\bm r}_i} 
[e^{i\delta({\bm k})} |i, d+\rangle + e^{-i\delta({\bm k})} |i, d-\rangle], \nonumber\\
i|{\bm k}, d_{y(\delta)z}\rangle\rangle &=&
{\cal N}^{-1/2} \sum_i e^{i{\bm k}\cdot{\bm r}_i} 
[e^{i\delta({\bm k})} |i, d+\rangle - e^{-i\delta({\bm k})} |i, d-\rangle],
\label{plane_waves}
\end{eqnarray}
where ${\cal N} = 2 N_{\rm Fe}$ is the number of iron site-orbitals.
The energy eigenvalue of the (bonding) $d_{x(\delta)z}$ band is given by
$\varepsilon_+({\bm k}) = \varepsilon_{\parallel}({\bm k}) + |\varepsilon_{\perp}({\bm k})|$ and
the energy eigenvalue of the (anti-bonding) $d_{y(\delta)z}$ band is given by
$\varepsilon_-({\bm k}) = \varepsilon_{\parallel}({\bm k}) - |\varepsilon_{\perp}({\bm k})|$,
where
%
\begin{subequations}
\begin{align}
\label{mtrx_lmnt_a}
\varepsilon_{\parallel}({\bm k}) &= -2 t_1^{\parallel} (\cos k_x a + \cos k_y a)
-2 t_2^{\parallel} (\cos k_+ a + \cos k_- a) \\
\label{mtrx_lmnt_b}
\varepsilon_{\perp}({\bm k}) &= -2 t_1^{\perp} (\cos k_x a - \cos k_y a)
-2 t_2^{\perp} (\cos k_+ a - \cos k_- a)
\end{align}
\end{subequations}
%
are diagonal and off-diagonal matrix elements,
with  $k_{\pm} = k_x \pm k_y$.
The phase shift $\delta({\bm k})$ is set by
$\varepsilon_{\perp}({\bm k}) = |\varepsilon_{\perp}({\bm k})| e^{i 2 \delta({\bm k})}$,
with
%
\begin{subequations}
\begin{align}
\label{c_2dlt}
\cos\,2\delta({\bm k}) &= {-t_1^{\perp}(\cos\, k_x a - \cos\, k_y a)\over
{\sqrt{t_1^{\perp 2}(\cos\, k_x a - \cos\, k_y a)^2 +
|2 t_2^{\perp}|^2 (\sin\, k_x a)^2 (\sin\, k_y a)^2}}}, \\
\label{s_2dlt}
\sin\,2\delta({\bm k}) &= {2 (t_2^{\perp} / i)(\sin\, k_x a) (\sin\, k_y a)\over
{\sqrt{t_1^{\perp 2}(\cos\, k_x a - \cos\, k_y a)^2 +
|2 t_2^{\perp}|^2 (\sin\, k_x a)^2 (\sin\, k_y a)^2}}}. 
\end{align}
\end{subequations}
%
At ${\bm k} = 0$ and ${\bm Q}_{\rm AF}$,
the matrix element $\varepsilon_{\perp}({\bm k})$ vanishes.
The phase factor $e^{i 2 \delta({\bm k})}$ is then notably singular there.

\begin{figure}
\includegraphics[scale=0.50, angle=0]{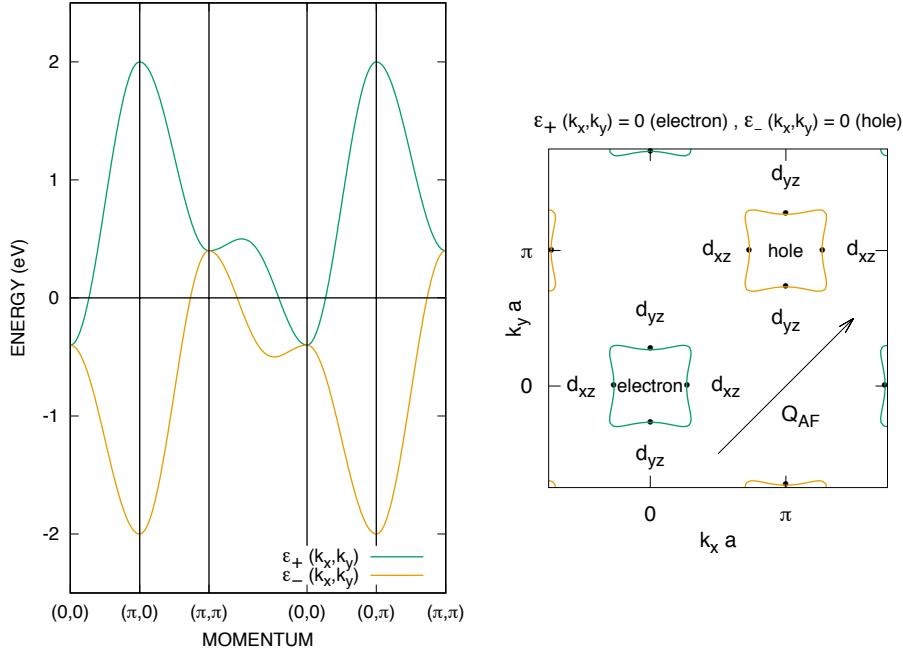}
\caption{Band structure with perfectly nested Fermi surfaces at half filling:
$\varepsilon_+({\bm k}) = 0$ and  $\varepsilon_-({\bm k}) = 0$,
with hopping matrix elements
$t_1^{\parallel} = 100$ meV, $t_1^{\perp} = 500$ meV, $t_2^{\parallel} = 0$,
and $t_2^{\perp} = 100\, i$ meV.
Dirac cones emerge from the dots on the Fermi surfaces in the hSDW state.}
\label{FS0}
\end{figure}

Now turn off next-nearest neighbor intra-orbital hopping: $t_2^{\parallel} = 0$.
The above energy bands then satisfy the perfect nesting condition\cite{jpr_rm_18}
\begin{equation}
\varepsilon_{\pm}({\bm k}+{\bm Q}_{\rm AF}) = - \varepsilon_{\mp}({\bm k}),
\label{prfct_nstng}
\end{equation}
where ${\bm Q}_{\rm AF} = (\pi/a,\pi/a)$ is the N\'eel ordering vector on the square
lattice of iron atoms. 
As a result, the Fermi level at half filling 
 lies at $\epsilon_{\rm F} = 0$.  
Figure \ref{FS0} shows such
perfectly nested electron-type and hole-type Fermi surfaces for hopping 
parameters $t_1^{\parallel} = 100$ meV, $t_1^{\perp} = 500$ meV, $t_2^{\parallel} = 0$
and $t_2^{\perp} = 100\, i$ meV.

\subsection{Extended Hubbard model}
Next, add interactions due to on-site Coulomb repulsion
and super-exchange interactions\cite{anderson_50,Si&A}
 via the Se atoms 
to the hopping Hamiltonian (\ref{hop}):
$H = H_{\rm hop} + H_U + H_{\rm sprx}$.
The second term counts on-site Coulomb repulsion\cite{2orb_Hbbrd},
\begin{eqnarray}
H_U =  \sum_i &&[U_0 n_{i,\alpha,\uparrow} n_{i,\alpha,\downarrow}
                +J_0 {\bm S}_{i, d-}\cdot {\bm S}_{i, d+} \nonumber \\
               &&+U_0^{\prime} n_{i,d+} n_{i,d-}
                +J_0^{\prime} (c_{i,d+,\uparrow}^{\dagger}c_{i,d+,\downarrow}^{\dagger}
                            c_{i,d-,\downarrow}c_{i,d-,\uparrow}+ {\rm h.c.})],
\label{U}
\end{eqnarray}
where $n_{i,\alpha,s} = c_{i,\alpha,s}^{\dagger}c_{i,\alpha,s}$ is the occupation operator, 
where ${\bm S}_{i,\alpha} = {\hbar\over 2}\sum_{s,s^{\prime}}
c_{i,\alpha,s}^{\dagger}{\boldsymbol \sigma}_{s,s^{\prime}} c_{i,\alpha,s^{\prime}}$
is the spin operator,
and where $n_{i,\alpha} = n_{i,\alpha,\uparrow} + n_{i,\alpha,\downarrow}$.
Above, $U_0>0$ denotes the intra-orbital on-site Coulomb repulsion energy,
while $U_0^{\prime} > 0$ denotes the inter-orbital one.
Also, $J_0 < 0$ is the ferromagnetic Hund's Rule exchange coupling constant,
while $J_0^{\prime}$ is the matrix element for on-site Josephson tunneling between orbitals.

The last term in the Hamiltonian represents 
super-exchange interactions\cite{anderson_50,Si&A}
among the iron spins via the selenium atoms:
\begin{eqnarray}
H_{\rm sprx} &=&
\sum_{\langle i,j \rangle}  J_1 ({\bm S}_{i, d-} + {\bm S}_{i, d+})
\cdot ({\bm S}_{j, d-} + {\bm S}_{j, d+}) \nonumber \\
&&+\sum_{\langle\langle  i,j \rangle\rangle} J_2 ({\bm S}_{i, d-} + {\bm S}_{i, d+})
\cdot ({\bm S}_{j, d-} + {\bm S}_{j, d+}).
\label{sprx}
\end{eqnarray}
Above, $J_1$ and $J_2$ are positive 
super-exchange coupling constants
across nearest neighbor and next-nearest neighbor iron sites.
Assume henceforth that magnetic frustration is moderate to strong:
$J_2 > 0.5 J_1$.
In isolation,
and at strong on-site-orbital repulsion,
$H_{\rm sprx}$ thereby favors
``stripe'' SDW order over conventional N\'eel order.

\section{Hidden Spin Density Wave}
The perfectly  nested Fermi surfaces shown by Fig. \ref{FS0} will result in 
a spin-density wave state
within the previous extended Hubbard model
at ordering wavevector ${\bm Q}_{\rm AF} = (\pi/a, \pi/a)$.  
In the present ${d-}/{d+}$ basis of $d_{xz}/d_{yz}$ orbitals,
the most natural candidates are the
true spin-density wave $(0,\pi,\pi)$
and the hidden spin-density wave $(\pi,\pi,\pi)$, 
with the ordered moments
\begin{equation}
{\cal N}^{-1}\sum_i e^{i {\bm Q}_{\rm AF}\cdot{\bm r}_i}
\sum_{s=\uparrow,\downarrow} ({\rm sgn}\, s)
{\hbar\over{2}} \langle c_{i,d+,s}^{\dagger} c_{i,d+,s} \pm c_{i,d-,s}^{\dagger} c_{i,d-,s}\rangle.
\label{hsdw_ordered_moment}
\end{equation}
It is important to recall that the creation/annihilation operators transform as
$$c_{i,d\pm,s}^{\dagger} \rightarrow e^{\pm i\phi} c_{i,d\pm,s}^{\dagger}\quad {\rm and} \quad
  c_{i,d\pm,s} \rightarrow e^{\mp i\phi} c_{i,d\pm,s}$$
under a rotation of the orbitals by an angle $\phi$ about the $z$ axis.
The ordered moments (\ref{hsdw_ordered_moment}) of both the true SDW ($+$)
and of the hidden SDW ($-$) are then notably isotropic with respect to such rotations.
Neither SDW state therefore couples to nematic instabilities that can appear in 
the phase diagram of iron superconductors \cite{xu_muller_sachdev_08,yoshizawa_simayi_12}.

\begin{table}
\begin{tabular}{|c|c|c|c|}
\hline
physical quantity & operator & $\ S\ $ & $\ I\ $ \\
\hline
density & $n_{i,d+} + n_{i,d-}$ & $0$ & $0$  \\
true spin & ${\bm S}_{i,d+} + {\bm S}_{i,d-}$ & $1$ & $0$ \\
hidden spin & ${\bm S}_{i,d+} - {\bm S}_{i,d-}$ & $1$ & $1$  \\
\hline
\end{tabular}
\caption{List of physical operators per site $i$
 according to spin and isospin quantum numbers, $S$ and $I$.
In the latter case,
the $d+$ and $d-$ orbitals are analogous to the $u$ and $d$ quarks.
(See ref. \cite{jpr_20b}.)}
\label{isospin}
\end{table}

Consider, now, the $d-$ and $d+$ orbitals as components of isospin\cite{jpr_20b} $I=1/2$.
In general, the ordered moment of an hSDW state has isospin $I=1$.
(See Table \ref{isospin}.)
In particular,
they are  components of the tensor product
\begin{equation}
 {\cal N}^{-1}
\sum_i e^{i {\bm Q}_{\rm AF}\cdot{\bm r}_i}
\sum_{s,s^{\prime}=\uparrow,\downarrow} {\boldsymbol \sigma}_{s,s^{\prime}}
\sum_{\alpha,\alpha^{\prime}=d-, d+} {\boldsymbol \tau}_{\alpha,\alpha^{\prime}}
{\hbar\over{2}}\langle c_{i,\alpha,s}^{\dagger} c_{i,\alpha^{\prime},s^{\prime}}\rangle,
\end{equation}
where ${\boldsymbol \sigma}$ and ${\boldsymbol \tau}$ are Pauli matrices.  
The candidate hSDW state (\ref{hsdw_ordered_moment}), for example,
 corresponds to the tensor product $\sigma_3\tau_3$.
Table \ref{hsw} lists the ordered magnetic moments of such hSDW states along
 the three principal axes of the isospin ${\bm I}$.
Notice the hSDW state along the $I_2$ isospin axis that was introduced by Berg, Metlitski and Sachdev
in the context of copper-oxide high-temperature superconductors \cite{BMS_12}. 
Both it and the hSDW state along the $I_1$ isospin axis are {\it not}, in fact,
isotropic about the orbital $z$ axis.
Below, we shall review the mean field theory for
 the candidate hSDW state\cite{jpr_rm_18} (\ref{hsdw_ordered_moment})
along the $I_3$ isospin axis.
Both it ($-$) and the true SDW state ($+$)
provide the basis for the RPA calculation in the next section.

\begin{table}
\begin{tabular}{|c|c|c|}
\hline
hSDW ordered moment & isospin axis & reference \\
\hline
$({\rm sgn}\, s) (c_{i,d_{xz},s}^{\dagger}  c_{i,d_{xz},s} - c_{i,d_{yz},s}^{\dagger} c_{i,d_{yz},s}) 
e^{i{\bm Q}_{\rm AF}\cdot {\bm r}_i}$ & $I_1$ & none \\
$({\rm sgn}\, s) (c_{i,d_{xz},s}^{\dagger} c_{i,d_{yz},s} + c_{i,d_{yz},s}^{\dagger} c_{i,d_{xz},s})
e^{i{\bm Q}_{\rm AF}\cdot {\bm r}_i}$ & $I_2$ & Berg, Metlitski and Sachdev (2012) \\
$({\rm sgn}\, s) i(c_{i,d_{xz},s}^{\dagger} c_{i,d_{yz},s} - c_{i,d_{yz},s}^{\dagger} c_{i,d_{xz},s})
e^{i{\bm Q}_{\rm AF}\cdot {\bm r}_i}$ & $I_3$ & Rodriguez (2017) \\
\hline
\end{tabular}
\caption{List of hidden-order magnetic moments by isospin quantization axis.
Examples of where such hidden SDW order parameters appear in the literature are also listed.}
\label{hsw}
\end{table}

\subsection{Mean Field Theory}\label{mft}
Assume that the expectation value of the magnetic moment per site, per orbital,
shows hidden N\'eel order, with spontaneous symmetry breaking along the $z$ axis: 
\begin{equation}
\langle m_{i,\alpha}\rangle = (-1)^{\alpha} e^{i{\bm Q}_{\rm AF}\cdot{\bm r}_i} \langle m_{0,0}\rangle,
\label{ordered_moment}
\end{equation}
where
$\langle m_{i,\alpha}\rangle = {1\over 2}
\langle n_{i,\alpha,\uparrow}\rangle-{1\over 2}\langle n_{i,\alpha,\downarrow}\rangle$.
(Henceforth, set $\hbar = 1$.)
Such an hSDW state ($-$) is expected to be more stable than the true SDW state ($+$)
in the presence of magnetic frustration\cite{jpr_10}, $J_1, J_2 > 0$. 
Also,
calculations in the local-moment limit find that
the above hidden magnetic order is more stable than the ``stripe'' SDW mentioned previously
at weak to moderate strength in the Hund's Rule coupling\cite{jpr_17,jpr_10}.
The super-exchange terms, $H_{\rm sprx}$, 
make no contribution within the mean-field approximation,
since the net magnetic moment per iron atom is null in the hidden-order N\'eel state.  
Also, the formation of a spin singlet per iron-site-orbital is suppressed
at the strong-coupling limit, $U_0\rightarrow\infty$.
The on-site-orbital Josephson tunneling term
($J_0^{\prime}$) in $H_U$ can then be neglected on that basis.
We are then left
with the two on-iron--site repulsion terms and the Hund's Rule term in $H_U$.

The mean-field replacement of the intra-orbital on-site term ($U_0$)
is the usual one\cite{hirsch_85}:
\begin{eqnarray}
n_{i,\alpha,\uparrow} n_{i,\alpha,\downarrow} \rightarrow 
{1\over 2}\langle n_{i,\alpha}\rangle (n_{i,\alpha,\uparrow}+n_{i,\alpha,\downarrow})
&-\langle m_{i,\alpha}\rangle (n_{i,\alpha,\uparrow} - n_{i,\alpha,\downarrow})\nonumber\\
&-\langle n_{i,\alpha,\uparrow}\rangle \langle n_{i,\alpha,\downarrow}\rangle.\nonumber
\end{eqnarray}
The first term above can be absorbed into the chemical potential
and
the last term above is a constant energy shift.
This  leaves a mean-field contribution to the Hamiltonian:
$-\sum_i\sum_{\alpha}  U_0\langle m_{i,\alpha}\rangle (n_{i,\alpha,\uparrow} - n_{i,\alpha,\downarrow})$.
A similar mean-field replacement of the inter-orbital on-iron-site repulsion term
($U_0^{\prime}$) in $H_U$
can be entirely absorbed into a shift of the chemical potential
plus  a constant energy shift\cite{jpr_rm_18}, on the other hand.  
Finally, we make the same type of mean-field replacement for the Hund's Rule
term ($J_0$) in $H_U$:
$${\bm S}_{i,d+} \cdot {\bm S}_{i,d-} \rightarrow 
 S_{i,d+}^{(z)} \langle S_{i,d-}^{(z)}\rangle 
+ \langle S_{i,d+}^{(z)}\rangle S_{i,d-}^{(z)}
- \langle S_{i,d+}^{(z)}\rangle \langle S_{i,d-}^{(z)}\rangle .$$
Again, the last term above is a constant energy shift.
The first two terms, however, contribute to the mean-field Hamiltonian:
$\sum_i \sum_{\alpha} {1\over 2} J_0 \langle m_{i,{\bar\alpha}}\rangle (n_{i,\alpha,\uparrow} - n_{i,\alpha,\downarrow})$,
which is equal to 
$-\sum_i \sum_{\alpha} {1\over 2} J_0 \langle m_{i,\alpha}\rangle (n_{i,\alpha,\uparrow} - n_{i,\alpha,\downarrow})$
in the case of hidden magnetic order (\ref{ordered_moment}).
Here, $\overline {d\pm} = d\mp$.

Neglecting on-site-orbital Josephson tunneling ($J_0^{\prime}$),
the net contribution to the mean-field Hamiltonian from interactions
in the present two-orbital Hubbard model is then
$$-\sum_i \sum_{\alpha} U(\pi) \langle m_{i,\alpha}\rangle (n_{i,\alpha,\uparrow} - n_{i,\alpha,\downarrow})
= - \langle m_{0,0} \rangle  U(\pi)
\sum_i \sum_{\alpha}
(-1)^{\alpha} e^{i{\bm Q}_{\rm AF}\cdot{\bm r}_i} 
(n_{i,\alpha,\uparrow} - n_{i,\alpha,\downarrow}),$$
where 
\begin{equation}
U(\pi) = U_0+{1\over 2} J_0.
\label{U(pi)}
\end{equation}
Notice that the sum on the right-hand side above over sites and over orbitals
 is twice the hidden-order moment
$S_z (\pi,{\bm Q}_{\rm AF})$. (See Appendix \ref{ppndx_m_e}.)
Re-expressing it in the band basis (\ref{plane_waves})
and then applying the identity (\ref{delta_k_pls_Q}) for the phase shift
ultimately yields the mean-field Hamiltonian for
the present two-orbital Hubbard model\cite{jpr_rm_18}:
\begin{eqnarray}
H^{(mf)} = && \sum_s\sum_{\bm k}\sum_{n=1,2}
 \varepsilon_n({\bm k}) c_s^{\dagger}(n,{\bm k}) c_s(n,{\bm k})  \nonumber \\
&& \mp \sum_s\sum_{\bm k}[({\rm sgn}\, s) \Delta({\bm k}) 
c_s^{\dagger}(1,{\bar{\bm k}}) c_s(2,{\bm k})+{\rm h.c.}],
\label{Hmf}
\end{eqnarray}
where ${\bar{\bm k}} = {\bm k}+{\bm Q}_{\rm AF}$,
with a gap function
\begin{equation}
\Delta({\bm k}) = \Delta_0 \sin[2\delta({\bm k})],
\label{gap}
\end{equation}
where
\begin{equation}
\Delta_0 = \langle m_{0,0}\rangle U(\pi).
\label{Delta0}
\end{equation}
Here, $c_s^{\dagger}(1,{\bm k})$ and $c_s^{\dagger}(2,{\bm k})$
create plane waves (\ref{plane_waves}) in the anti-bonding ($d_{y(\delta)z}$)
and bonding ($d_{x(\delta)z}$) bands, respectively.
Here also,
intra-band scattering
has been neglected because it shows no nesting.
After shifting the momentum of the  anti-bonding band ($n=1$)
by ${\bm Q}_{AF}$,
we arrive at the final form of the mean-field Hamiltonian:
\begin{eqnarray}
H^{(mf)} = & \sum_s\sum_{\bm k}\varepsilon_+({\bm k}) [c_s^{\dagger}(2,{\bm k}) c_s(2,{\bm k})
- c_s^{\dagger}(1,{\bar{\bm k}}) c_s(1,{\bar{\bm k}})]  \nonumber \\
&  \mp  \sum_s\sum_{\bm k}[({\rm sgn}\, s)\Delta({\bm k}) 
c_s^{\dagger}(1,{\bar{\bm k}}) c_s(2,{\bm k})+{\rm h.c.}].
\label{HMF}
\end{eqnarray}

For convenience,
now set the $\pm$ sign that originates from the orbital matrix elements to minus.
[See Appendix \ref{ppndx_m_e} and (\ref{delta_k_pls_Q}).]
The mean-field Hamiltonian (\ref{HMF}) is diagonalized 
in the usual way by writing the electron in terms of quasi-particle 
excitations\cite{schrieffer_wen_zhang_89,singh_tesanovic_90,chubokov_frenkel_92}:
\begin{eqnarray}
c_s^{\dagger}(2,{\bm k}) &=& u({\bm k}) \alpha_s^{\dagger}(2,{\bm k})
 - ({\rm sgn}\, s) v({\bm k}) \alpha_s^{\dagger}(1,{\bar{\bm k}}) , \nonumber\\
c_s^{\dagger}(1,{\bar{\bm k}}) &=& ({\rm sgn}\, s) v({\bm k})  \alpha_s^{\dagger}(2,{\bm k}) 
 + u({\bm k}) \alpha_s^{\dagger}(1,{\bar{\bm k}}) .
\label{qps}
\end{eqnarray}
Here, $u({\bm k})$ and $v({\bm k})$ are coherence factors with square magnitudes
\begin{equation}
u^2 = {1\over 2} + {1\over 2} {\varepsilon_+\over E} \quad{\rm and}\quad
v^2 = {1\over 2} - {1\over 2} {\varepsilon_+\over E},
\label{u&v}
\end{equation}
where $E({\bm k}) = [\varepsilon_+^2({\bm k}) + \Delta^2({\bm k})]^{1/2}$.
The mean-field Hamiltonian can then be expressed in terms of the occupation
of quasiparticles as
\begin{equation}
H^{(mf)} = \sum_s\sum_{\bm k} E({\bm k})[\alpha_s^{\dagger}(2,{\bm k}) \alpha_s(2,{\bm k}) 
- \alpha_s^{\dagger}(1,{\bar{\bm k}}) \alpha_s (1,{\bar{\bm k}})].
\label{dgnl}
\end{equation}
The quasi-particle excitation energies are then $E({\bm k})$ for particles
and $E({\bar {\bm k}})$ for holes,
with a gap function (\ref{gap}) that has $D_{xy}$ symmetry.
Dirac nodes therefore emerge from the points on the Fermi surfaces indicated
by Fig. \ref{FS0}.
At half filling, the energy band $-E({\bar {\bm k}})$ is filled
and the energy band $+E({\bm k})$ is empty.
Last, inverting (\ref{qps}) yields
\begin{eqnarray}
\alpha_s^{\dagger}(2,{\bm k}) &=& u({\bm k}) c_s^{\dagger}(2,{\bm k})
 + ({\rm sgn}\, s) v({\bm k}) c_s^{\dagger}(1,{\bar{\bm k}}) , \nonumber\\
\alpha_s^{\dagger}(1,{\bar{\bm k}}) &=& - ({\rm sgn}\, s)v({\bm k})  c_s^{\dagger}(2,{\bm k}) 
 + u({\bm k}) c_s^{\dagger}(1,{\bar{\bm k}}) .
\label{QPS}
\end{eqnarray}
Quasiparticles are a coherent superposition of an electron
of momentum ${\bm k}$ in the bonding ($+$) band $2$
with an electron of momentum ${\bm k}+{\bm Q}_{\rm AF}$ 
in the anti-bonding ($-$) band $1$.

\begin{figure}
\includegraphics[scale=0.50]{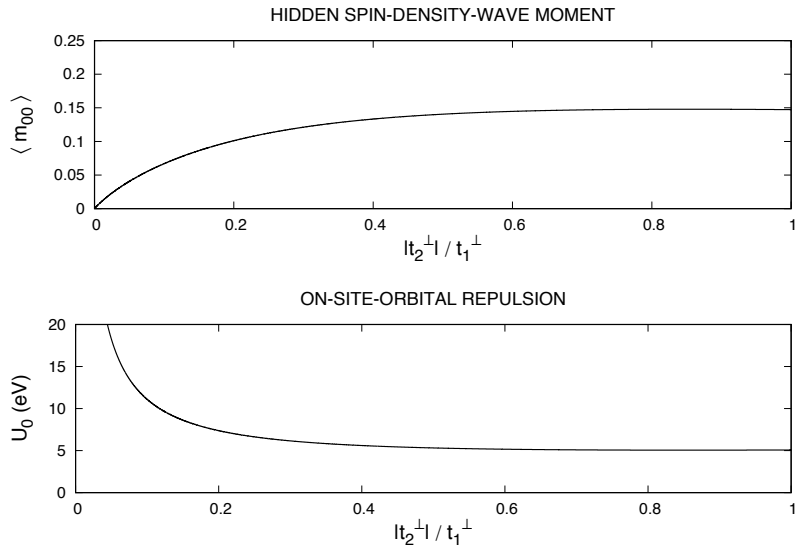}
\caption{Displayed is the ordered magnetic moment obtained from
 the gap equation at $\Delta_0 = 740$ meV versus $t_2^{\perp}/i$,
 along with the corresponding on-site-orbital Hubbard $U_0$.
The Hund's Rule coupling is fixed at $J_0 = - 100$ meV.
Also, the square lattice of iron atoms is a periodic $1000\times 1000$ grid.
The remaining  electron hopping parameters are listed in the caption to Fig. \ref{FS0}.}
\label{hidden_moment}
\end{figure}

Finally, to obtain the gap equation, 
we exploit the pattern of hidden N\'eel order
(\ref{ordered_moment}), 
and write the gap maximum (\ref{Delta0}) as
$$\Delta_0 = 
{\cal N}^{-1}\sum_i\sum_{\alpha} U(\pi) \langle m_{i,\alpha}\rangle (-1)^{\alpha} e^{i{\bm Q}_{\rm AF}\cdot {\bm r}_i} 
= {\cal N}^{-1} U(\pi) \langle S_z(\pi,{\bm Q}_{\rm AF})\rangle.$$
Using expressions for the hidden-order moment in terms of band states
yields
$$\Delta_0 = -
{\cal N}^{-1}{1\over 2}\sum_s\sum_{\bm k}\sum_{n} U(\pi) ({\rm sgn}\, s) [\sin\, 2\delta({\bm k})]
\langle c_s^{\dagger}({\bar n},{\bar{\bm k}}) c_s(n,{\bm k})\rangle,$$
where ${\bar n} = 1 + (n\; {\rm mod}\; 2)$.
[See Appendix \ref{ppndx_m_e} and (\ref{delta_k_pls_Q}).]
Intra-band scattering has again been neglected.
Substituting in (\ref{qps}) and the conjugate annihilation operators,
and recalling that the $n=1$ quasi-particle band is filled, 
while the $n=2$ quasi-particle band is empty,
yields 
$\langle c_s^{\dagger}({\bar n},{\bar{\bm k}}) c_s(n,{\bm k})\rangle
 = - ({\rm sgn}\, s) u({\bm k}) v({\bm k})$
for the expectation value.
We thereby obtain
$$\Delta_0 = 
{\cal N}^{-1}\sum_{\bm k} U(\pi)
[\sin\, 2\delta({\bm k})] \Delta({\bm k})/E({\bm k}),$$
or equivalently, the gap equation
\begin{equation}
1 = U(\pi) {\cal N}^{-1} \sum_{\bm k} 
{[\sin\, 2\delta({\bm k})]^2 \over{\sqrt{\varepsilon_+^2({\bm k})+\Delta_0^2 [\sin\, 2\delta({\bm k})]^2}}}.
\label{gap_eq}
\end{equation}
Figure \ref{hidden_moment} displays solutions of the gap equation at constant $\Delta_0$.
It is important to mention that they depend only on the hopping parameters and on $U(\pi)$.
By (\ref{U(pi)}), $\Delta_0$ then is also constant along a line, $U_0$ versus $-J_0$,
such that $U(\pi)$ remains constant.

%


\subsection{Lifshitz Transition of the Fermi Surfaces}
\begin{figure}
\includegraphics[scale=0.50, angle=0]{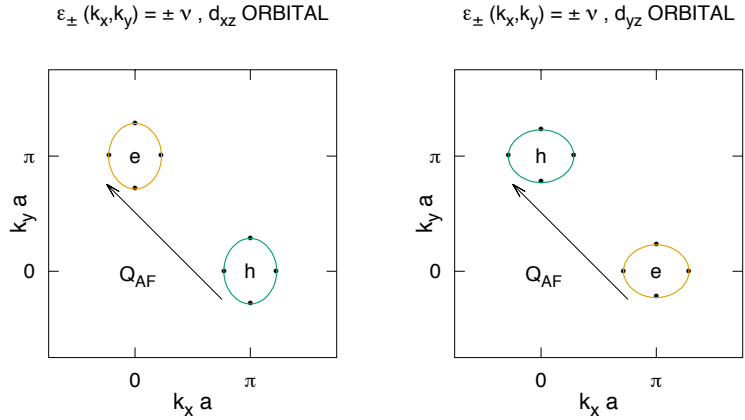}
\caption{
Shown are the renormalized Fermi surface pockets after a Lifshitz transition due to
fluctuation-exchange with hidden spinwaves centered at ${\bm Q}_{\rm AF}$.
[See ref. \cite{jpr_rm_18} and Eq. (\ref{chi_44}).]
The orbital character is only approximate,
although it becomes exact as the area of the Fermi surface pockets vanishes as $U_0$ diverges.
Dirac cones emerge from the dots on the renormalized Fermi surfaces
 in the fluctuation-corrected hSDW state.}
\label{FS1}
\end{figure}

Before going on to the calculation of the dynamical spin susceptibility of
 the hSDW state within RPA in the next section,
 it is important to point out that
ARPES on electron-doped iron-selenide high-$T_c$ superconductors
generally sees only electron-type Fermi surface pockets
 at the corner of the folded (two-iron) Brillouin zone\cite{zhou_13,peng_14,lee_14,zhao_16}.
The perfectly nested Fermi surfaces displayed by Fig. \ref{FS0} do {\it not},
therefore, coincide with ARPES measurements on these materials.
The following RPA of the extended Hubbard model for electron-doped iron selenide
reveals hidden spinwaves (\ref{chi_44})
 that disperse acoustically from
the antiferromagnetic wavevector, ${\bm Q}_{\rm AF}$, however.
 In the critical hSDW state,
as  $\Delta_0 \rightarrow 0$,
the author and a co-worker have recently shown that
fluctuation-exchange interactions of the electrons
 with such Goldstone modes
result in a Lifshitz transition of the nested Fermi surfaces
displayed by\cite{jpr_rm_18} Fig. \ref{FS0}:
the electron-type band $\varepsilon_+({\bm k})$ is pulled down in energy
with respect to the Fermi level, while the hole-type band $\varepsilon_-({\bm k})$ is pulled
up in energy by an equal and opposite amount. The Lifshitz transition results in
electron/hole pockets near the opposite band edges at moderate to large
Hubbard repulsion $U_0$.  Figure \ref{FS1} displays the resulting renormalized
Fermi surfaces for hopping parameters that are listed in the caption to Fig. \ref{FS0}.
Also, the above Lifshitz transition is accompanied by wavefunction renormalizations
that result in vanishingly small quasi-particle weight at the renormalized Fermi levels\cite{jpr_rm_18}.

The Lifshitz transition described above was predicted
 at half filling for the critical hSDW state
($\Delta_0 \rightarrow 0$)
via an Eliashberg-type theory of hidden spin-fluctuation exchange
 in the particle-hole channel\cite{jpr_rm_18}.
The critical hSDW itself can be achieved by tuning the strength of  Hund's Rule
to the transition point where a true SDW state appears.
Adding electrons above half filling
 suggests a rigid shift up in energy of the Fermi level 
with respect to the renormalized band structure,
$\varepsilon_+({\bm k}) - \nu$ and $\varepsilon_-({\bm k})+\nu$.
Here, $\nu > 0$ represents the energy shift due to the Lifshitz transition.
It lies just below the upper band edge of the bonding ($+$) band.
At saturation, a rigid shift in energy of such a renormalized band structure
 results in Fermi surface points for the new hole-type Fermi surfaces shown in Fig. \ref{FS1},
and in new electron-type Fermi surface pockets that are a bit larger than those shown in Fig. \ref{FS1}.
Such a rigid energy shift has in fact been confirmed by the author
in a related Eliashberg theory for hidden spin-fluctuation exchange,
but in the conventional particle-particle channel\cite{jpr_20b}.
In particular, the author finds that the quasi-particle weight of the holes
remains vanishingly small at the Fermi level, 
while that the quasi-particle weight of the electrons can be appreciable at the Fermi level.
This scenario is confirmed by a local-moment model for the present extended Hubbard model
that harbors the hSDW state\cite{jpr_17}.

In the following section,
we will proceed to compute the dynamical spin susceptibility of the hSDW within RPA,
but starting from the unrenormalized electron bands shown in Fig. \ref{FS0}.
This is justified on the basis of perturbation theory in powers of the interactions,
 $H_U$ and $H_{\rm sprx}$.
Does that conflict with the Lifshitz transition\cite{jpr_rm_18} shown by Fig. \ref{FS1}?
We believe that is does not.
By (\ref{prfct_nstng}),
the renormalized energy bands mentioned above trivially also  satisfy perfect nesting:
\begin{eqnarray}
\varepsilon_{+}({\bm k}+{\bm Q}_{\rm AF}) - \nu = - [\varepsilon_{-}({\bm k}) + \nu],\nonumber \\
\varepsilon_{-}({\bm k}+{\bm Q}_{\rm AF}) + \nu = - [\varepsilon_{+}({\bm k}) - \nu].
\label{gnrl_prfct_nstng}
\end{eqnarray}
The {\it form} of the RPA to be developed below does  not therefore change if the shifted energy
bands are used instead, along with the wavefunction renormalization.  We therefore believe that
starting the RPA below from the unrenormalized bands (Fig. \ref{FS0})  is compatible with the
Lifshitz transition\cite{jpr_rm_18} mentioned above.

\section{Spin Fluctuations within Random Phase Approximation}
Is the previous mean-field solution for
the hSDW state of the extended Hubbard model
 for electron-doped iron selenide\cite{jpr_rm_18} stable?
To answer this question,
we shall compute the transverse dynamical spin susceptibility
within the random phase approximation.
Like in the original ``spin bag'' calculation of the SDW state
in the conventional Hubbard model over the square lattice\cite{schrieffer_wen_zhang_89},
the bare dynamical spin susceptibilities (RPA bubbles)
do not conserve crystal moment over the square lattice,
whereas the interaction terms do.
In the present case, additionally, the bare RPA bubbles also break
orbital-swap symmetry, $P_{d,{\bar d}}$, because of orbital mixing ($t_2^{\perp}$),
while the interaction terms preserve that symmetry as well.

\subsection{Bare Spin Fluctuations at Perfect Nesting}
We shall first compute the {\it bare} spin-fluctuation propagators in the hSDW state,
at perfect nesting (\ref{prfct_nstng}).
Recall the spin-flip operator at relative momentum ${\bm q}$,
in the true or hidden channel, $q_0 = 0$ or $\pi$:
\begin{eqnarray}
S^+ (q_0,{\bm q}) = 
\sum_i\sum_{\alpha=0,1} e^{i q_0 \alpha} e^{i{\bm q}\cdot{\bm r}_i}
c_{i,\alpha,\uparrow}^{\dagger} c_{i,\alpha,\downarrow}.
\label{SS}
\end{eqnarray}
Here, the indices $\alpha = 0$ and $1$ represent the $d-$ and $d+$ orbitals, respectively.
In the band basis set by
 the plane-wave eigenstates (\ref{plane_waves}) of the hopping Hamiltonian,
it has the form
\begin{equation}
S^+ (q_0,{\bm q}) = \sum_{\bm k}\sum_{n,n^{\prime}}
{\cal M}^{(q_0)}_{n,{\bm k};n^{\prime},{\bm k}^{\prime}} \, c_{\uparrow}^{\dagger}(n^{\prime},{\bm k}^{\prime})
c_{\downarrow} (n,{\bm k}),
\label{ss}
\end{equation}
with ${\bm k}^{\prime} = {\bm k} + {\bm q}$.
Above, the indices $n = 1$ and $2$ represent the anti-bonding and bonding bands
that are in momentum-dependent orbitals
$(-i) d_{y(\delta)z}$ and $d_{x(\delta)z}$, respectively.
The orbital matrix element is computed in Appendix \ref{ppndx_m_e}, 
and it is given by
\begin{equation}
{\cal M}_{n,{\bm k};n^{\prime},{\bm k}^{\prime}}^{(m \pi)} =
\begin{cases}
\cos[\delta({\bm k})-\delta({\bm k}^{\prime})] & {\rm for}\quad n^{\prime} = n + m\; ({\rm mod}\; 2),\\
-i\, \sin[\delta({\bm k})-\delta({\bm k}^{\prime})] & {\rm for}\quad n^{\prime} = n + m + 1\; ({\rm mod}\; 2).
\end{cases}
\label{A_MM}
\end{equation}
%

Now define the Nambu-Gorkov spinor that incorporates the physics of nesting\cite{jpr_rm_18}:
\begin{equation}
C_s({\bm k}) =
\left[ {\begin{array}{c}
c_{s}(2,{\bm k}) \\ c_{s}(1,{\bar{\bm k}}).
\end{array} } \right].
\label{spinor}
\end{equation}
The spin-flip operator (\ref{ss}) can then be broken up into four components by the
$2\times 2$ identity matrix, $\tau_0$, 
and by the Pauli matrices, $\tau_1$, $\tau_2$, and $\tau_3$:
\begin{equation}
S_{\mu}^+ (q_0,{\bm q}) = \sum_{\bm k}
{\cal M}^{(q_0)}_{{\bm k},{\bm k}^{\prime}} (\mu) \, C_{\uparrow}^{\dagger}({\bm k}^{\prime}) \tau_{\mu}
C_{\downarrow} ({\bm k}),
\label{Ss}
\end{equation}
with matrix elements
\begin{equation}
{\cal M}^{(0)}_{{\bm k},{\bm k}^{\prime}} (\mu) =
\begin{cases}
{\cal M}^{(0)}_{1,{\bar{\bm k}};1,{\bar{\bm k}}^{\prime}} = {\cos}[\delta({\bm k})-\delta({\bm k}^{\prime})] = {\cal M}^{(0)}_{2,{\bm k};2,{\bm k}^{\prime}}\quad & {\rm if}\ \mu = 0\ {\rm (not\ nested)},\\
i{\cal M}^{(0)}_{1,{\bar{\bm k}};2,{\bm k}^{\prime}} = \pm \cos[\delta({\bm k})+\delta({\bm k}^{\prime})] = -i{\cal M}^{(0)}_{2,{\bm k};1,{\bar{\bm k}}^{\prime}}\quad & {\rm if}\ \mu = 2\ {\rm (nested)} \\
0 & {\rm if} \ \mu = 1, 3,
\end{cases}
\label{M_0}
\end{equation}
and
\begin{equation}
{\cal M}^{(\pi)}_{{\bm k},{\bm k}^{\prime}} (\mu) =
\begin{cases}
-{\cal M}^{(\pi)}_{1,{\bar{\bm k}};1,{\bar{\bm k}}^{\prime}} = -i \sin[\delta({\bm k})-\delta({\bm k}^{\prime})] = {\cal M}^{(\pi)}_{2,{\bm k};2,{\bm k}^{\prime}}\quad & {\rm if}\ \mu = 3\ {\rm (not\ nested)},\\
{\cal M}^{(\pi)}_{1,{\bar{\bm k}};2,{\bm k}^{\prime}} = \pm\sin[\delta({\bm k})+\delta({\bm k}^{\prime})] = {\cal M}^{(\pi)}_{2,{\bm k};1,{\bar{\bm k}}^{\prime}}\quad & {\rm if}\ \mu = 1\ {\rm (nested)},\\
0 & {\rm if} \ \mu = 0, 2.
\end{cases}
\label{M_pi}
\end{equation}
Here, we have used the property
\begin{equation}
\delta({\bm k} + {\bm Q}_{\rm AF}) = \pm {\pi\over 2} - \delta({\bm k})
\label{delta_k_pls_Q}
\end{equation}
satisfied by the phase shift, which is a result of the property
$\varepsilon_{\perp}({\bm k} + {\bm Q}_{\rm AF}) = - \varepsilon_{\perp}^*({\bm k})$
satisfied by the matrix element (\ref{mtrx_lmnt_b}).
The components $S_{\mu}^+ (q_0,{\bm q})$ of the spin-flip operator (\ref{SS})
can then be re-assembled following the nesting (1) versus the non-nesting (0) nature
of the momentum transfer, ${\bm q}$:
\begin{eqnarray}
S_{q_0,0}^{+}({\bm q}) &= S_{0}^+(q_0,{\bm q}) + S_{3}^+(q_0,{\bm q})\quad & {(\rm not\ nested)} ,\nonumber \\
S_{q_0,1}^{+}({\bm q}) &= S_{1}^+(q_0,{\bm q}) + S_{2}^+(q_0,{\bm q})\quad & {(\rm nested)} .
\label{s0_s1}
\end{eqnarray}
Inspection of (\ref{M_0}) and (\ref{M_pi}) then yields that the above spin operators take the form
\begin{equation}
S_{q_0,\gamma}^{+}({\bm q}) = \sum_{\bm k} {\cal M}_{{\bm k},{\bm k}^{\prime}}^{(q_0,\gamma)}
C_{\uparrow}^{\dagger}({\bm k}^{\prime}) \tau_{(q_0,\gamma)} C_{\downarrow} ({\bm k}) ,
\label{S_q0_gmm}
\end{equation}
where the products ${\cal M}_{{\bm k},{\bm k}^{\prime}}^{(q_0,\gamma)} \tau_{(q_0,\gamma)}$
are listed in Table \ref{M_tau}.

\begin{table}
\begin{tabular}{|c|c|c|}
\hline
                    & not nested ($0$) & nested ($1$) \\
\hline
true spin ($0$)     & $\cos[\delta({\bm k})-\delta({\bm k}^{\prime})]\, \tau_0$  & $\pm\, \cos[\delta({\bm k})+\delta({\bm k}^{\prime})]\, \tau_2$ \\
\hline
hidden spin ($\pi$) & $-i\, \sin[\delta({\bm k})-\delta({\bm k}^{\prime})]\, \tau_3$  & $\pm\, \sin[\delta({\bm k})+\delta({\bm k}^{\prime})]\, \tau_1$ \\
\hline
\end{tabular}
\caption{The products ${\cal M}_{{\bm k},{\bm k}^{\prime}}^{(q_0,\gamma)} \, \tau_{(q_0\gamma)}$
that appear in $S_{q_0,\gamma}^{+}({\bm q})$, 
where $q_0 = 0, \pi$ are labels for true versus hidden spin,
and where $\gamma = 0,1$ are labels for un-nested versus nested momentum transfer.
[See Eqs. (\ref{ss})-(\ref{S_q0_gmm}).]}
\label{M_tau}
\end{table}

Next,
en route to computing the bare spin-fluctuation propagator
of the hSDW state within mean field theory,
we will first compute the Nambu-Gorkov Greens function.
Let $C_s({\bm k},t)$ denote the time evolution of 
the destruction operators (\ref{spinor}) $C_s({\bm k})$,
and let $C_s^{\dagger}({\bm k},t)$ denote the time evolution
of the conjugate creation operators $C_s^{\dagger}({\bm k})$.
The Nambu-Gorkov electron propagator is then
the Fourier transform 
$i G_s({\bm k},\omega) = \int d t_{1,2} e^{i \omega t_{1,2}}
\langle T[C_s({\bm k},t_1) C_s^{\dagger}({\bm k},t_2)]\rangle$,
where $t_{1,2} = t_1 - t_2$, and where $T$ is the time-ordering operator.
It is a $2 \times 2$  matrix. 
By expression (\ref{HMF}) for the mean-field Hamiltonian,
the matrix inverse of the Nambu-Gorkov Greens function takes the form
\begin{equation}
G_s^{-1}({\bm k},\omega) =
  \omega\, \tau_0 
- \varepsilon_+({\bm k})\, \tau_3
\pm ({\rm sgn}\, s) \Delta({\bm k})\,\tau_1.
\label{1/G}
\end{equation}
%
Here,
$\Delta({\bm k})$ is the quasi-particle gap (\ref{gap}).
Notice that the term proportional to $\tau_3$
 is a direct consequence   of perfect nesting  (\ref{prfct_nstng}).
Matrix inversion of (\ref{1/G}) yields 
the Nambu-Gorkov Greens function\cite{jpr_rm_18,nambu_60,gorkov_58,schrieffer_64}
$G = \sum_{\mu = 0}^{3} G^{(\mu)} \tau_{\mu}$, with components
\begin{eqnarray}
G_s^{(0)} &=& {1\over{2}}
\Biggl({1\over{\omega-E}} + {1\over{\omega+E}}\Biggr),\nonumber \\
G_s^{(1)} &=& \mp {1\over{2}}
\Biggl({1\over{\omega-E}} - {1\over{\omega+E}}\Biggr)
{\Delta\over E} ({\rm sgn}\, s), \nonumber \\
G_s^{(2)} &=& 0, \nonumber \\
G_s^{(3)} &=& {1\over{2}}
\Biggl({1\over{\omega-E}} - {1\over{\omega+E}}\Biggr)
{\varepsilon_+\over  E}.
\label{G}
\end{eqnarray}
%
Above, the excitation energy is
$E = (\varepsilon_+^2 + \Delta^2)^{1/2}$.

We shall now define the bare dynamical spin susceptibility of the hSDW state
 with indices composed of true/hidden spin ($q_0$)
and of un-nested/nested momentum transfer ($\gamma$):
\begin{equation}
\chi_{q_0,\gamma;q_0^{\prime},\gamma^{\prime}}^{(0) + -} ({\bm q},\omega) = {i\over{\cal N}}
\langle S_{q_0,\gamma}^{+} ({\bm q},\omega) S_{q_0^{\prime},\gamma^{\prime}}^{-} ({\bm q},\omega) \rangle .
\end{equation}
Here, $S_{q_0,\gamma}^{+} ({\bm q},\omega)$ is the Fourier transform 
of the time-evolution of the spin-flip operator, $S_{{q_0},\gamma}^{+} ({\bm q})$.
[See (\ref{s0_s1}) and (\ref{S_q0_gmm}).]
Analytically continuing this dynamical spin susceptibility to imaginary time yields
a  convolution in terms of Matsubara frequencies:
\begin{equation}
\chi_{q_0,\gamma;q_0^{\prime},\gamma^{\prime}}^{(0) + -} ({\bm q},i\omega_m)= {k_B T\over{\cal N}} \sum_{i\omega_n} \sum_{\bm k}
{\rm tr}[
G_{\uparrow} ({\bm k}^{\prime},i\omega_{n^{\prime}}) 
 \tau_{(q_0,\gamma)} 
G_{\downarrow}({\bm k},i\omega_n)
 \tau_{(q_0^{\prime},\gamma^{\prime})}]
{\cal M}_{{\bm k},{\bm k^{\prime}}}^{(q_0,\gamma)}
{\cal M}_{{\bm k},{\bm k^{\prime}}}^{(q_0^{\prime},\gamma^{\prime}) *} ,
\label{Chi}
\end{equation}
where the orbital matrix element
${\cal M}_{{\bm k},{\bm k}^{\prime}}^{(q_0,\gamma)}$
appears as a product with the $2\times 2$ matrix $\tau_{(q_0,\gamma)}$
in Table \ref{M_tau}.
Here, ${\bm k}^{\prime} = {\bm k} + {\bm q}$
and $i\omega_{n^{\prime}} = i\omega_n + i\omega_m$.
Substituting in the Nambu-Gorkov Greens function (\ref{G})
yields the expression
\begin{equation}
\chi_{p_0,\gamma;q_0,\delta}^{(0) + -} ({\bm q},i\omega_m) = {k_B T\over{\cal N}} \sum_{i\omega_n} \sum_{\bm k}
\sum_{\mu,\nu = 0}^3
{\rm tr}[\tau_{\mu} \tau_{(p_0,\gamma)} \tau_{\nu} \tau_{(q_0,\delta)}]
G_{\uparrow}^{(\mu)} ({\bm k}^{\prime},i\omega_{n^{\prime}})
G_{\downarrow}^{(\nu)}({\bm k},i\omega_n)
{\cal M}_{{\bm k},{\bm k}^{\prime}}^{(p_0,\gamma)}
{\cal M}_{{\bm k},{\bm k}^{\prime}}^{(q_0,\delta) *} 
\label{chi}
\end{equation}
for the bare dynamical spin susceptibility.

It is well known that
 the sum over Matsubara frequencies in the expression above
for the bare dynamical spin susceptibility (\ref{chi})
 can be evaluated in terms of Fermi-Dirac distribution functions.
  Below, we obtain the corresponding Lindhard functions
in the zero-temperature limit.   The required trace formulas for products of
$2\times 2$ matrices are listed in Appendix \ref{ppndx_traces}.

\begin{enumerate}
\item $(0,0;0,0)$: true spin; true spin\\
${\cal M}_{{\bm k},{\bm k}^{\prime}}^{(0,0)}
{\cal M}_{{\bm k},{\bm k}^{\prime}}^{(0,0) *}  =
\cos^2 (\delta-\delta^{\prime})$ and
${\rm tr}(\tau_{\mu} \tau_{0} \tau_{\nu} \tau_{0}) = 2 \delta_{\mu,\nu}$.
Then
$$\sum_{\mu,\nu = 0}^3 
{\rm tr}(\tau_{\mu} \tau_{0} \tau_{\nu} \tau_{0}) G_{\uparrow}^{\prime (\mu)} G_{\downarrow}^{(\nu)} =
2[G_{\uparrow}^{\prime (0)} G_{\downarrow}^{(0)} +
G_{\uparrow}^{\prime (1)} G_{\downarrow}^{(1)} +
G_{\uparrow}^{\prime (3)} G_{\downarrow}^{(3)}].$$
Hence,
\begin{eqnarray}
\chi_{0,0;0,0}^{(0) + -} ({\bm q},\omega) = {1\over{2 {\cal N}}} &\sum_{\bm k}
& \biggl(1-{\varepsilon_+ \varepsilon_+^{\prime} - \Delta \Delta^{\prime}\over{E E^{\prime}}}\biggr)
 \biggl({1\over{E + E^{\prime} - \omega}} + 
{1\over{E + E^{\prime} + \omega}}\biggr) \nonumber \\
&& \cdot {1\over 2} [1 + (\cos 2\delta) (\cos 2\delta^{\prime}) + (\sin 2\delta) (\sin 2\delta^{\prime})].
\label{chi_0000}
\end{eqnarray}

\item $(0,0;\pi,0)$: true spin; hidden spin\\
${\cal M}_{{\bm k},{\bm k}^{\prime}}^{(0,0)}
{\cal M}_{{\bm k},{\bm k}^{\prime}}^{(\pi,0) *}  =
i \cos (\delta-\delta^{\prime}) \sin (\delta-\delta^{\prime})$ and
${\rm tr}(\tau_{\mu} \tau_{0} \tau_{\nu} \tau_{3}) = 
2 (\delta_{\mu,0} \delta_{\nu,3} + \delta_{\mu,3} \delta_{\nu,0}+ i\, \epsilon_{\mu,\nu,3})$, \\
where $\epsilon_{\mu,\nu,i}$ coincides with the Levi-Civita tensor for $\mu,\nu = 1, 2, 3$,
while it vanishes otherwise, for $\mu = 0$, or for $\nu = 0$.
Then
$$\sum_{\mu,\nu = 0}^3 
{\rm tr}(\tau_{\mu} \tau_{0} \tau_{\nu} \tau_{3}) G_{\uparrow}^{\prime (\mu)} G_{\downarrow}^{(\nu)} =
2[G_{\uparrow}^{\prime (0)} G_{\downarrow}^{(3)} +
G_{\uparrow}^{\prime (3)} G_{\downarrow}^{(0)}].$$
Hence,
\begin{eqnarray}
\chi_{0,0;\pi,0}^{(0) + -} ({\bm q},\omega) = - {1\over{2 {\cal N}}} &\sum_{\bm k}
& \biggl({\varepsilon_+ \over{E }} - {\varepsilon_+^{\prime}\over{E^{\prime}}}\biggr)
 \biggl({1\over{E + E^{\prime} - \omega}} - 
{1\over{E + E^{\prime} + \omega}}\biggr) \nonumber \\
&& \cdot {i\over 2} [(\sin 2\delta) (\cos 2\delta^{\prime}) - (\cos 2\delta) (\sin 2\delta^{\prime})].
\label{chi_00p0}
\end{eqnarray}
%

\item $(0,0;0,1)$: true spin; SDW moment\\
${\cal M}_{{\bm k},{\bm k}^{\prime}}^{(0,0)}
{\cal M}_{{\bm k},{\bm k}^{\prime}}^{(0,1) *}  = \pm
\cos (\delta-\delta^{\prime}) \cos (\delta+\delta^{\prime})$ and
${\rm tr}(\tau_{\mu} \tau_{0} \tau_{\nu} \tau_{2}) = 
2 (\delta_{\mu,0} \delta_{\nu,2} + \delta_{\mu,2}\delta_{\nu,0} + i\, \epsilon_{\mu,\nu,2})$.
Then
$$\sum_{\mu,\nu = 0}^3 
{\rm tr}(\tau_{\mu} \tau_{0} \tau_{\nu} \tau_{3}) G_{\uparrow}^{\prime (\mu)} G_{\downarrow}^{(\nu)} =
2 i[G_{\uparrow}^{\prime (3)} G_{\downarrow}^{(1)} -
G_{\uparrow}^{\prime (1)} G_{\downarrow}^{(3)}].$$
Hence,
\begin{eqnarray}
\chi_{0,0;0,1}^{(0) + -} ({\bm q},\omega) = - {1\over{2 {\cal N}}} &\sum_{\bm k}
& {\varepsilon_+ \Delta^{\prime} + \Delta \varepsilon_+^{\prime}\over{ E E^{\prime}}}
 \biggl({1\over{E + E^{\prime} - \omega}} +
{1\over{E + E^{\prime} + \omega}}\biggr) \nonumber \\
&& \cdot {i\over 2} [(\cos 2\delta) + (\cos 2\delta^{\prime})].
\label{chi_0001}
\end{eqnarray}

\item $(0,0;\pi,1)$: true spin; hSDW moment\\
${\cal M}_{{\bm k},{\bm k}^{\prime}}^{(0,0)}
{\cal M}_{{\bm k},{\bm k}^{\prime}}^{(\pi,1) *}  = \pm
\cos (\delta-\delta^{\prime}) \sin (\delta+\delta^{\prime})$ and
${\rm tr}(\tau_{\mu} \tau_{0} \tau_{\nu} \tau_{1}) = 
2 (\delta_{\mu,0} \delta_{\nu,1} + \delta_{\mu,1} \delta_{\nu,0} + i\, \epsilon_{\mu,\nu,1})$. \\
Then
$$\sum_{\mu,\nu = 0}^3 
{\rm tr}(\tau_{\mu} \tau_{0} \tau_{\nu} \tau_{1}) G_{\uparrow}^{\prime (\mu)} G_{\downarrow}^{(\nu)} =
2[G_{\uparrow}^{\prime (0)} G_{\downarrow}^{(1)} +
G_{\uparrow}^{\prime (1)} G_{\downarrow}^{(0)}].$$
Hence,
\begin{eqnarray}
\chi_{0,0;\pi,1}^{(0) + -} ({\bm q},\omega) = - {1\over{2 {\cal N}}} &\sum_{\bm k}
& \biggl({\Delta\over{E }}+{\Delta^{\prime}\over{E^{\prime}}}\biggr)
 \biggl({1\over{E + E^{\prime} - \omega}} - 
{1\over{E + E^{\prime} + \omega}}\biggr) \nonumber \\
&& \cdot {1\over 2} [(\sin 2\delta) + (\sin 2\delta^{\prime})].
\label{chi_00p1}
\end{eqnarray}

\item $(\pi,0;\pi,0)$: hidden spin; hidden spin\\
${\cal M}_{{\bm k},{\bm k}^{\prime}}^{(\pi,0)}
{\cal M}_{{\bm k},{\bm k}^{\prime}}^{(\pi,0) *}  =
\sin^2 (\delta-\delta^{\prime})$ and
${\rm tr}(\tau_{\mu} \tau_{3} \tau_{\nu} \tau_{3}) = 2\, {\rm sgn}_{\mu} (3) \delta_{\mu,\nu}$, \\
where ${\rm sgn}_{\mu} (i) = 1$ if $\mu = 0$ or $i$,
and where ${\rm sgn}_{\mu} (i) = -1$ otherwise.
Then
$$\sum_{\mu,\nu = 0}^3
{\rm tr}(\tau_{\mu} \tau_{3} \tau_{\nu} \tau_{3}) G_{\uparrow}^{\prime (\mu)} G_{\downarrow}^{(\nu)} =
2[G_{\uparrow}^{\prime (0)} G_{\downarrow}^{(0)} -
G_{\uparrow}^{\prime (1)} G_{\downarrow}^{(1)} +
G_{\uparrow}^{\prime (3)} G_{\downarrow}^{(3)}].$$
Hence,
\begin{eqnarray}
\chi_{\pi,0;\pi,0}^{(0) + -} ({\bm q},\omega) = {1\over{2 {\cal N}}} &\sum_{\bm k}
& \biggl(1-{\varepsilon_+ \varepsilon_+^{\prime} + \Delta \Delta^{\prime}\over{E E^{\prime}}}\biggr)
 \biggl({1\over{E + E^{\prime} - \omega}} + 
{1\over{E + E^{\prime} + \omega}}\biggr) \nonumber \\
&& \cdot {1\over 2} [1 - (\cos 2\delta) (\cos 2\delta^{\prime}) - (\sin 2\delta) (\sin 2\delta^{\prime})].
\label{chi_p0p0}
\end{eqnarray}

\item $(\pi,0;0,1)$: hidden spin; SDW moment\\
${\cal M}_{{\bm k},{\bm k}^{\prime}}^{(\pi,0)}
{\cal M}_{{\bm k},{\bm k}^{\prime}}^{(0,1) *}  = \mp
 i \sin (\delta-\delta^{\prime}) \cos (\delta+\delta^{\prime})$ and \\
${\rm tr}(\tau_{\mu} \tau_{3} \tau_{\nu} \tau_{2}) =
2 (\delta_{\mu,3} \delta_{\nu,2} + \delta_{\mu,2} \delta_{\nu,3}
+ i\, \delta_{\mu,0} \epsilon_{3,\nu,2} + i\, \delta_{\nu,0} \epsilon_{\mu,3,2})$.
Then
$$\sum_{\mu,\nu = 0}^3 
{\rm tr}(\tau_{\mu} \tau_{3} \tau_{\nu} \tau_{2}) G_{\uparrow}^{\prime (\mu)} G_{\downarrow}^{(\nu)} =
2 i[G_{\uparrow}^{\prime (0)} G_{\downarrow}^{(1)} -
G_{\uparrow}^{\prime (1)} G_{\downarrow}^{(0)}].$$
Hence,
\begin{eqnarray}
\chi_{\pi,0;0,1}^{(0) + -} ({\bm q},\omega) = - {1\over{2 {\cal N}}} &\sum_{\bm k}
& \biggl({\Delta \over{E }} - {\Delta^{\prime}\over{E^{\prime}}}\biggr)
 \biggl({1\over{E + E^{\prime} - \omega}} -
{1\over{E + E^{\prime} + \omega}}\biggr) \nonumber \\
&& \cdot {1\over 2} [(\sin 2\delta) - (\sin 2\delta^{\prime})].
\label{chi_p001}
\end{eqnarray}
%

\item $(\pi,0;\pi,1)$: hidden spin; hSDW moment\\
${\cal M}_{{\bm k},{\bm k}^{\prime}}^{(\pi,0)}
{\cal M}_{{\bm k},{\bm k}^{\prime}}^{(\pi,1) *}  = \mp
 i \sin (\delta-\delta^{\prime}) \sin (\delta+\delta^{\prime})$ and \\
${\rm tr}(\tau_{\mu} \tau_{3} \tau_{\nu} \tau_{1}) =
2 (\delta_{\mu,3} \delta_{\nu,1} + \delta_{\mu,1} \delta_{\nu,3}
+ i\, \delta_{\mu,0} \epsilon_{3,\nu,1} + i\, \delta_{\nu,0} \epsilon_{\mu,3,1})$.
Then
$$\sum_{\mu,\nu = 0}^3 
{\rm tr}(\tau_{\mu} \tau_{3} \tau_{\nu} \tau_{1}) G_{\uparrow}^{\prime (\mu)} G_{\downarrow}^{(\nu)} =
2[G_{\uparrow}^{\prime (3)} G_{\downarrow}^{(1)} +
G_{\uparrow}^{\prime (1)} G_{\downarrow}^{(3)}].$$
Hence,
\begin{eqnarray}
\chi_{\pi,0;\pi,1}^{(0) + -} ({\bm q},\omega) =   {1\over{2 {\cal N}}} &\sum_{\bm k}
& {\varepsilon_+ \Delta^{\prime}  - \Delta \varepsilon_+^{\prime}\over{E E^{\prime}}}
 \biggl({1\over{E + E^{\prime} - \omega}} +
{1\over{E + E^{\prime} + \omega}}\biggr) \nonumber \\
&& \cdot {i\over 2} [(\cos 2\delta) - (\cos 2\delta^{\prime})].
\label{chi_p0p1}
\end{eqnarray}

\item $(0,1;0,1)$: SDW moment; SDW moment\\
${\cal M}_{{\bm k},{\bm k}^{\prime}}^{(0,1)}
{\cal M}_{{\bm k},{\bm k}^{\prime}}^{(0,1) *}  =
\cos^2 (\delta+\delta^{\prime})$ and
${\rm tr}(\tau_{\mu} \tau_{2} \tau_{\nu} \tau_{2}) = 2\, {\rm sgn}_{\mu} (2)  \delta_{\mu,\nu}$.
Then
$$\sum_{\mu,\nu = 0}^3
{\rm tr}(\tau_{\mu} \tau_{2} \tau_{\nu} \tau_{2}) G_{\uparrow}^{\prime (\mu)} G_{\downarrow}^{(\nu)} =
2[G_{\uparrow}^{\prime (0)} G_{\downarrow}^{(0)} -
G_{\uparrow}^{\prime (1)} G_{\downarrow}^{(1)} -
G_{\uparrow}^{\prime (3)} G_{\downarrow}^{(3)}].$$
Hence,
\begin{eqnarray}
\chi_{0,1;0,1}^{(0) + -} ({\bm q},\omega) = {1\over{2 {\cal N}}} &\sum_{\bm k}
& \biggl(1+{\varepsilon_+ \varepsilon_+^{\prime} - \Delta \Delta^{\prime}\over{E E^{\prime}}}\biggr)
 \biggl({1\over{E + E^{\prime} - \omega}} + 
{1\over{E + E^{\prime} + \omega}}\biggr) \nonumber \\
&& \cdot {1\over 2} [1 + (\cos 2\delta) (\cos 2\delta^{\prime}) - (\sin 2\delta) (\sin 2\delta^{\prime})].
\label{chi_0101}
\end{eqnarray}

\item $(0,1;\pi,1)$: SDW moment; hSDW moment\\
${\cal M}_{{\bm k},{\bm k}^{\prime}}^{(0,1)}
{\cal M}_{{\bm k},{\bm k}^{\prime}}^{(\pi,1) *}  =
\cos (\delta+\delta^{\prime}) \sin (\delta+\delta^{\prime})$ and \\
${\rm tr}(\tau_{\mu} \tau_{2} \tau_{\nu} \tau_{1}) =
2 (\delta_{\mu,2} \delta_{\nu,1} + \delta_{\mu,1} \delta_{\nu,2}
+ i\, \delta_{\mu,0} \epsilon_{2,\nu,1} + i\, \delta_{\nu,0} \epsilon_{\mu,2,1})$.
Then
$$\sum_{\mu,\nu = 0}^3 
{\rm tr}(\tau_{\mu} \tau_{2} \tau_{\nu} \tau_{1}) G_{\uparrow}^{\prime (\mu)} G_{\downarrow}^{(\nu)} =
2 i[G_{\uparrow}^{\prime (0)} G_{\downarrow}^{(3)} -
G_{\uparrow}^{\prime (3)} G_{\downarrow}^{(0)}].$$
Hence,
\begin{eqnarray}
\chi_{0,1;\pi,1}^{(0) + -} ({\bm q},\omega) = - {1\over{2 {\cal N}}} &\sum_{\bm k}
& \biggl({\varepsilon_+\over{E}} + {\varepsilon_+^{\prime}\over{E^{\prime}}}\biggr)
 \biggl({1\over{E + E^{\prime} - \omega}} -
{1\over{E + E^{\prime} + \omega}}\biggr) \nonumber \\
&& \cdot {i\over 2} [(\sin 2\delta)(\cos 2\delta^{\prime}) + (\cos 2\delta)(\sin 2\delta^{\prime})].
\label{chi_01p1}
\end{eqnarray}

\item $(\pi,1;\pi,1)$: hSDW moment; hSDW moment\\
${\cal M}_{{\bm k},{\bm k}^{\prime}}^{(\pi,1)}
{\cal M}_{{\bm k},{\bm k}^{\prime}}^{(\pi,1) *}  =
\sin^2 (\delta+\delta^{\prime})$ and
${\rm tr}(\tau_{\mu} \tau_{1} \tau_{\nu} \tau_{1}) = 2\, {\rm sgn}_{\mu} (1) \delta_{\mu,\nu}$. \\
Then
$$\sum_{\mu,\nu = 0}^3
{\rm tr}(\tau_{\mu} \tau_{1} \tau_{\nu} \tau_{1}) G_{\uparrow}^{\prime (\mu)} G_{\downarrow}^{(\nu)} =
2[G_{\uparrow}^{\prime (0)} G_{\downarrow}^{(0)} +
G_{\uparrow}^{\prime (1)} G_{\downarrow}^{(1)} -
G_{\uparrow}^{\prime (3)} G_{\downarrow}^{(3)}].$$
Hence,
\begin{eqnarray}
\chi_{\pi,1;\pi,1}^{(0) + -} ({\bm q},\omega) = {1\over{2 {\cal N}}} &\sum_{\bm k}
& \biggl(1+{\varepsilon_+ \varepsilon_+^{\prime} + \Delta \Delta^{\prime}\over{E E^{\prime}}}\biggr)
 \biggl({1\over{E + E^{\prime} - \omega}} + 
{1\over{E + E^{\prime} + \omega}}\biggr) \nonumber \\
&& \cdot {1\over 2} [1 - (\cos 2\delta) (\cos 2\delta^{\prime}) + (\sin 2\delta) (\sin 2\delta^{\prime})].
\label{chi_p1p1}
\end{eqnarray}

\end{enumerate}
Last, inspection of the trace formulas for products of $2\times 2$ matrices
listed in Appendix \ref{ppndx_traces} yields
that the matrix formed by the trace
${\rm tr}(\tau_{\mu} \tau_{\gamma} \tau_{\nu} \tau_{\delta})$
as a function of
the indices $\gamma$ and $\delta$ is hermitian.
The matrix of bare spin susceptibilities is then also hermitian by expression (\ref{chi}).
The remaining off-diagonal bare spin susceptibilities are then complex conjugates
of those listed above.

\subsection{Random Phase Approximation}
Next, to construct the RPA, we must determine how the interaction terms in
$H_U$ (\ref{U}) and in $H_{\rm sprx}$ (\ref{sprx})
couple to the previous bare dynamical spin susceptibilities.
All of the interaction terms are translationally invariant.  
They are also all invariant under orbital swap, $P_{d,{\bar d}}$: $d\pm \rightarrow d\mp$.
Both momentum and parity, ${\bm q}$ and $q_0$, are then good quantum numbers for
the interactions $H_U$ and $H_{\rm sprx}$.  
They are therefore diagonal in momentum and in parity.
Yet what are such diagonal matrix elements of $H_U$ and of $H_{\rm sprx}$?

\begin{figure}
\includegraphics[scale=0.50]{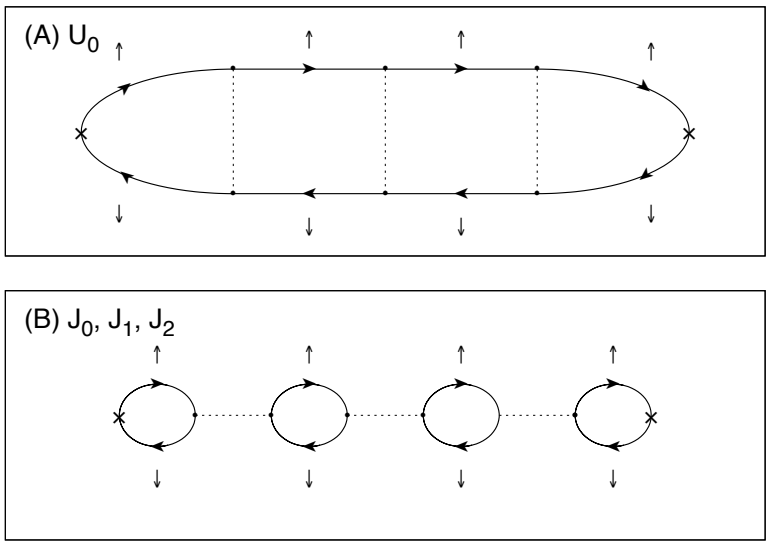}
\caption{Representative Feynman diagrams for the dynamical transverse  spin susceptibility,
$\chi^{+ -}({\bm q},\omega)$, of the extended two-orbital Hubbard model within RPA.}
\label{feynman}
\end{figure}

The on-site-orbital Hubbard repulsion  ($U_0$) has the form
$n_{\uparrow} n_{\downarrow} = + c_{\uparrow}^{\dagger} c_{\downarrow}^{\dagger} c_{\downarrow} c_{\uparrow}$.
On the other hand,
the spin-flip part of the on-site Hund's Rule coupling ($J_0$)
and of the super-exchange interactions ($J_{1}$ and $J_{2}$)
have the transverse Heisenberg-exchange form
$${1\over 2} S^{+} S^{\prime -} + {1\over 2} S^{-} S^{\prime +} =
-{1\over 2} c_{\uparrow}^{\dagger} c_{\downarrow}^{\prime\dagger} c_{\downarrow} c_{\uparrow}^{\prime}
-{1\over 2} c_{\downarrow}^{\dagger} c_{\uparrow}^{\prime\dagger} c_{\uparrow} c_{\downarrow}^{\prime}.$$
Figure \ref{feynman} displays the corresponding Feynman diagrams for the RPA.
Taking the Fourier transform of the previous interactions  in site-orbital space yields
the following spin-flip contribution to the interaction:
\begin{eqnarray}
V^{+-}(q_0,{\bm q}) & = &
 U_0 - {1\over 2} J_0 \cos (q_0) \nonumber \\
&& - \delta_{q_0,0}\{2 J_1[\cos (q_x a) + \cos (q_y a)] + 2 J_2 [\cos (q_+ a) + \cos(q_- a)]\}.
\label{V+-}
\end{eqnarray}
Here, $q_{\pm} = q_x \pm q_y$.
Last, inter-orbital on-site interactions ($U_0^{\prime}$) can be neglected because they couple only to density,
while on-site Josephson tunneling ($J_0^{\prime}$) can be neglected at strong on-site repulsion $U_0$.

The true-spin and the hidden-spin components of the spin-flip potential $V^{+-}(q_0,{\bm q})$ 
are listed in the first-two rows of Table \ref{v+-}.
They clearly scatter  fermions at un-nested momentum transfer, ${\bm q}$ small.  
The last two rows in Table \ref{v+-},
however, are the corresponding spin-flip interaction potentials 
that scatter fermions at momentum transfer that is indeed nested, ${\bm q}$ small.
These are simply shifted with respect to the former un-nested spin-flip potentials
by the antiferromagnetic wavevector  ${\bm Q}_{\rm AF} = (\pi/a, \pi/a)$.
Adding up the Dyson series of Feynman diagrams of the types displayed by Fig. \ref{feynman}
yields the RPA for the dynamical spin susceptibility:
\begin{equation}
\chi^{+-}({\bm q},\omega) = \chi^{(0)+-}({\bm q},\omega) [1 - V^{+-}({\bm q}) \chi^{(0)+-}({\bm q},\omega)]^{-1} .
\label{RPA}
\end{equation}
Above,
$V^{+-}({\bm q})$ is a $4\times 4$ matrix with diagonal matrix elements that are listed in Table \ref{v+-},
and with off-diagonal matrix elements that are null.
The matrix elements of the bare dynamical spin susceptibility, $\chi^{(0)+-}({\bm q},\omega)$,
 are listed above in the previous subsection.

\begin{table}
\begin{tabular}{|c|c|}
\hline
 $(q_0,\gamma)$ & $V_{q_0,\gamma}^{+-}({\bm q})$  \\
\hline
true spin $(0, 0)$ & $U_0 - {1\over 2}J_0 - 2 J_1[\cos (q_x a) + \cos (q_y a)] - 2 J_2 [\cos (q_+ a) + \cos(q_- a)]$ \\ 
\hline
hidden spin $(\pi, 0)$ & $U_0 + {1\over 2}J_0$  \\
\hline
SDW moment $(0, 1)$ & $U_0 - {1\over 2}J_0 + 2 J_1[\cos (q_x a) + \cos (q_y a)] - 2 J_2 [\cos (q_+ a) + \cos(q_- a)]$ \\
\hline
hSDW moment $(\pi, 1)$ & $U_0 + {1\over 2} J_0$ \\
\hline
\end{tabular}
\caption{Interactions in momentum space per true ($0$) and hidden ($\pi$) spin quantum numbers.
The SDW and hSWD interactions are the nested versions of the previous;
i.e., ${\bm q}\rightarrow {\bm q}+{\bm Q}_{\rm AF}$.}
\label{v+-}
\end{table}
%


\subsection{Reflection Symmetries and the Long Wavelength Limit}
\begin{figure}
\includegraphics[scale=0.50]{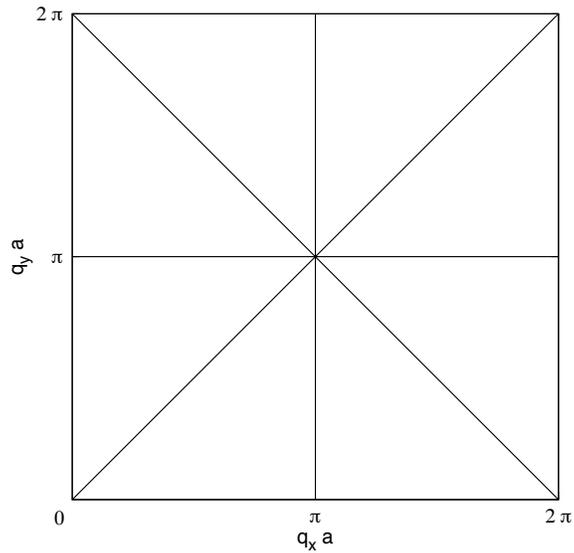}
\caption{Principal  axes in the first Brillouin zone.}
\label{prncpl_xs}
\end{figure}

In general, the bare dynamical spin susceptibility, $\chi^{(0)+-}({\bm q},\omega)$,
is a dense $4\times 4$ matrix.  It and the RPA solution (\ref{RPA})
break down into block-diagonal $2\times 2$ matrices at momentum transfers that are
along a principal axis of the first Brillouin zone, however.
To demonstrate this, suppose that the momentum transfer ${\bm q}$
lies ({\it i})
along one of the horizontal or vertical principal axes
 of the Brillouin zone shown by Fig. \ref{prncpl_xs}.
Reflections about such principal axes act on momenta as
\begin{eqnarray}
R_{x}: && (k_x,k_y) \rightarrow (k_x,-k_y), \nonumber \\
R_{y}: && (k_x,k_y) \rightarrow (-k_x,k_y).
\end{eqnarray}
Inspection of expressions (\ref{c_2dlt})  and (\ref{s_2dlt})
then yields that the components of the orbital phase factor transform under such reflections as
\begin{equation}
R_{x(y)}: (\cos 2\delta, \sin 2\delta) \rightarrow (\cos 2\delta, -\sin 2\delta).
\end{equation}
Next, suppose instead that the momentum transfer ${\bm q}$
lies ({\it ii}) along one of the diagonal principal axes
 of the Brillouin zone shown by Fig. \ref{prncpl_xs}.
Reflections about such principal axes act on momenta as
\begin{eqnarray}
R_{x^{\prime}}: && (k_x,k_y) \rightarrow (k_y,k_x), \nonumber \\
R_{y^{\prime}}: && (k_x,k_y) \rightarrow (-k_y,-k_x),
\end{eqnarray}
on the other hand.
Inspection of expressions (\ref{c_2dlt})  and (\ref{s_2dlt})
then yields that the components of the orbital phase factor transform under such reflections as
\begin{equation}
R_{x^{\prime}(y^{\prime})}: (\cos 2\delta, \sin 2\delta) \rightarrow (-\cos 2\delta, \sin 2\delta).
\end{equation}
Observe, now, that the energy eigenvalue $\varepsilon_+({\bm k})$ is invariant under all such 
reflections about a principal axis.
Inspection of the integrands of the bare dynamical spin susceptibilities,
$\chi_{p_0,\gamma;q_0,\delta}^{(0)+-}({\bm q},\omega)$,
listed above then yields unique parities under all such reflections
for ${\bm q}$ along the reflection axis.
They are listed in Table \ref{parities}.
We thereby conclude that the off-diagonal components
of the bare dynamical spin susceptibility with negative parities 
are {\it null} for momentum transfer ${\bm q}$ along a principal axis.

At momentum transfer ${\bm q}$ along a principal axis,
the RPA solution (\ref{RPA}) for the dynamical spin susceptibility therefore decouples 
into two $2\times 2$ blocks among the pairs of components $(1,4)$ and $(2,3)$:
\begin{equation}
\begin{bmatrix}
\chi_{11}^{+-} & \chi_{14}^{+-} \\
\chi_{41}^{+-} & \chi_{44}^{+-}
\end{bmatrix}
= {1\over{d(1,4)}}
\begin{bmatrix}
\chi_{11}^{(0)+-} & \chi_{14}^{(0)+-} \\
\chi_{41}^{(0)+-} & \chi_{44}^{(0)+-} 
\end{bmatrix}
\begin{bmatrix}
1-V_4^{+-}\chi_{44}^{(0)+-} & +V_1^{+-}\chi_{14}^{(0)+-} \\
+V_4^{+-}\chi_{41}^{(0)+-} & 1-V_{1}^{+-}\chi_{11}^{(0)+-} 
\end{bmatrix},
\label{14}
\end{equation}
with determinant
\begin{equation}
d(1,4) =(1-V_1^{+-}\chi_{11}^{(0)+-})(1-V_4^{+-}\chi_{44}^{(0)+-}) - 
V_1^{+-} V_4^{+-} \chi_{14}^{(0)+-}\chi_{41}^{(0)+-},
\label{d14}
\end{equation}
and
\begin{equation}
\begin{bmatrix}
\chi_{22}^{+-} & \chi_{23}^{+-} \\
\chi_{32}^{+-} & \chi_{33}^{+-}
\end{bmatrix}
= {1\over{d(2,3)}}
\begin{bmatrix}
\chi_{22}^{(0)+-} & \chi_{23}^{(0)+-} \\
\chi_{32}^{(0)+-} & \chi_{33}^{(0)+-} 
\end{bmatrix}
\begin{bmatrix}
1-V_3^{+-}\chi_{33}^{(0)+-} & +V_2^{+-}\chi_{23}^{(0)+-} \\
+V_3^{+-}\chi_{32}^{(0)+-} & 1-V_{2}^{+-}\chi_{22}^{(0)+-} 
\end{bmatrix},
\label{23}
\end{equation}
with determinant
\begin{equation}
d(2,3) =(1-V_2^{+-}\chi_{22}^{(0)+-})(1-V_3^{+-}\chi_{33}^{(0)+-}) - 
V_2^{+-} V_3^{+-} \chi_{23}^{(0)+-}\chi_{32}^{(0)+-}.
\label{d23}
\end{equation}
Above, we have enumerated the indices
for true spin, for hidden spin, for the SDW moment, and for the hSDW moment
 by 
$1 = (0,0)$,
$2 = (\pi,0)$,
$3 = (0,1)$, and
$4 = (\pi,1)$.
These results will be evaluated numerically at low frequency in the next section.

\begin{table}
\begin{tabular}{|c|c|c|c|c|}
\hline
  $R$      &   $0, 0$   &  $\pi, 0$  &  $0, 1$   &  $\pi, 1$   \\
\hline
  $0, 0$   &     $+$    &    $-$     &    $-$    &     $+$     \\
\hline
  $\pi, 0$ &     $-$    &    $+$     &    $+$    &     $-$     \\
\hline
  $0, 1$   &     $-$    &    $+$     &    $+$    &     $-$     \\
\hline
  $\pi, 1$ &     $+$    &    $-$     &    $-$    &     $+$     \\
\hline
\end{tabular}
\caption{Parities of the integrands of the bare dynamical spin susceptibility,
$\chi_{p_0,\gamma;q_0,\delta}^{(0)+-}({\bm q},\omega)$,
under reflection, $R$, about a principal axis of the Brillouin zone,
at momentum transfers, ${\bm q}$, along the same axis.
(See Fig. \ref{prncpl_xs}.)}
\label{parities}
\end{table}

Let us first, however, apply the previous to
 reveal the Goldstone modes associated
with hidden magnetic order (\ref{ordered_moment}).
Consider then the determinant (\ref{d14}) that describes the dynamics
of the principal hidden antiferromagnetic order parameter
 at small momentum transfer along the $x$ axis: ${\bm q} = (q_x,0)$.
The factor $1 - V_4^{+-}({\bm q}) \chi_{44}^{(0)+-}({\bm q},\omega)$ vanishes at
${\bm q} =0$ and $\omega = 0$ because of the gap equation (\ref{gap_eq}).
After expanding the determinant (\ref{d14}) to lowest non-trivial order in $q_x$ and in $\omega$,
we then get
\begin{eqnarray}
d(1,4) &\cong& [1-U(0)\chi_{\perp}^{(0)}] 
 \Bigl[-\omega^2 {U(\pi)\over{(2\Delta_0)^2}} \chi_{\perp}^{(0)} +
    q_x^2 {U(\pi)\over{(2\Delta_0)^2}} \rho_s\Bigr] - 
 \omega^2 {U(0) U(\pi)\over{(2\Delta_0)^2}} \chi_{\perp}^{(0) 2} \nonumber \\
&\cong&-\omega^2 {U(\pi)\over{(2\Delta_0)^2}} \chi_{\perp}^{(0)} +
 [1-U(0)\chi_{\perp}^{(0)}] 
    q_x^2 {U(\pi)\over{(2\Delta_0)^2}} \rho_s ,
\end{eqnarray}
where
\begin{equation}
\chi_{\perp}^{(0)} = {1\over{\cal N}} \sum_{\bm k} {\Delta_0^2 (\sin 2\delta)^2\over{E^3}}
\label{chi0_perp}
\end{equation}
is the bare transverse spin susceptibility\cite{jpr_rm_18},
and where $\rho_s$ denotes the spin rigidity of the hSDW state.
Here, $U(0) = U_0 - {1\over 2} J_0 -4 J_1 -4 J_2$
and $U(\pi) = U_0 + {1\over 2} J_0$.
Setting the determinant to zero, $d(1,4) = 0$, then yields an acoustic dispersion
for the Goldstone modes associated with hidden magnetic order,
$\omega = c_0 |{\bm q}|$, with a hidden spin-wave velocity,
$c_0 = (\rho_s / \chi_{\perp})^{1/2}$,
set by the spin rigidity, $\rho_s$,
and by the transverse spin susceptibility within RPA,
\begin{equation}
\chi_{\perp} = \chi_{\perp}^{(0)} / [1 - U(0) \chi_{\perp}^{(0)}].
\label{chi_perp}
\end{equation}
The former acoustic spectrum for hidden spinwaves
 will be computed numerically in the next section.
(See Fig. \ref{spctrm}.)
Also, substituting in the lowest-order values
$\chi_{11}^{(0)+-} \cong \chi_{\perp}^{(0)}$ and
$\chi_{44}^{(0)+-} \cong 1/U(\pi)$
for  the matrix elements of the bare spin susceptibility into the RPA expression (\ref{14})
yields the dynamical spin susceptibility for hidden spin waves
at long wavelength and low frequency:
\begin{equation}
\chi_{44}^{+-}({\bm q},\omega) \cong {(2\langle m_{0,0}\rangle)^2\over{\chi_{\perp}}}
{1\over{c_0^2|{\bm q}|^2 - \omega^2}},
\label{chi_44}
\end{equation}
where $\langle m_{0,0}\rangle$ is the ordered moment for the hSDW state (\ref{ordered_moment}).
We thereby recover the result expected from hydrodynamics
for the dynamical correlation function of
 the  antiferromagnetic ordered moment\cite{halperin_hohenberg_69,forster_75}.

\section{Numerical Evaluation of RPA}
Below, we reveal the spin excitations of the hSDW state
within the extended Hubbard model for electron-doped iron selenide
over a periodic square lattice of iron atoms.
Specifically,
the dynamical spin susceptibility is evaluated numerically at half filling within RPA.

\subsection{Propagation along  Principal Axes at Low Frequency}
Let us again suppose that the momentum ${\bm q}$ carried by a spin excitation in the hSDW state
lies along one of the principal axes displayed by Fig. \ref{prncpl_xs}.
It was demonstrated at the end of the previous section
that the RPA solution (\ref{RPA}) decouples 
into dynamics between the true spin and the primary hSDW order parameter $(1,4)$,
and into dynamics between the hidden spin and the secondary SDW order parameter $(2,3)$.
Equations (\ref{14}) and (\ref{23}), specifically,
give the respective dynamical spin susceptibilities within RPA.
In order to obtain the low-energy spectrum of such spin excitations,
we can next expand the bare spin susceptibilities to lowest non-trivial order in frequency.
In the case of the dynamics of the primary order parameter, for example, we have
\begin{eqnarray}
\chi_{11}^{(0)+-}({\bm q},\omega) &\cong & \chi_{11}^{(0)}({\bm q}) + \omega^2 \chi_{11}^{(2)}({\bm q}),\\
\chi_{14}^{(0)+-}({\bm q},\omega) &\cong & \omega \chi_{14}^{(1)}({\bm q}),\\
\chi_{44}^{(0)+-}({\bm q},\omega) &\cong & \chi_{44}^{(0)}({\bm q}) + \omega^2 \chi_{44}^{(2)}({\bm q}),
\end{eqnarray}
where $\chi_{11}^{(0)}({\bm q}) =\chi_{11}^{(0)+-}({\bm q},0)$,
where $\chi_{44}^{(0)}({\bm q}) =\chi_{44}^{(0)+-}({\bm q},0)$,
where
\begin{eqnarray}
\chi_{11}^{(2)}({\bm q}) = {1\over{\cal N}} 
&\sum_{\bm k}& \biggl(1-{\varepsilon_+\varepsilon_+^{\prime}-\Delta\Delta^{\prime}\over{E E^{\prime}}}\biggr)
{1\over{(E+E^{\prime})^3}} \nonumber \\
&& \cdot {1\over 2}[1+(\cos 2\delta)(\cos 2\delta^{\prime})+(\sin 2\delta)(\sin 2\delta^{\prime})],\\
\chi_{44}^{(2)}({\bm q}) = {1\over{\cal N}} 
&\sum_{\bm k}& \biggl(1+{\varepsilon_+\varepsilon_+^{\prime}+\Delta\Delta^{\prime}\over{E E^{\prime}}}\biggr)
{1\over{(E+E^{\prime})^3}} \nonumber \\
&& \cdot {1\over 2}[1-(\cos 2\delta)(\cos 2\delta^{\prime})+(\sin 2\delta)(\sin 2\delta^{\prime})],
\end{eqnarray}
and  where
\begin{equation}
\chi_{14}^{(1)}({\bm q}) = - {1\over{\cal N}}
\sum_{\bm k} \biggl({\Delta\over{E}}+{\Delta^{\prime}\over{E^{\prime}}}\biggr){1\over 2}
{\sin(2\delta) + \sin(2\delta^{\prime})\over{(E+E^{\prime})^2}}.
\end{equation}
Recall that we have enumerated the indices for the true spin and for the hidden SDW moment by
$1 = (0,0)$ and by
$4 = (\pi,1)$, respectively.
The RPA denominator (\ref{d14}) then has the form
$d(1,4) = P - \omega^2 Q$,
where $P$ and $Q$ are functions of momentum transfer ${\bm q}$ that are given by
\begin{eqnarray}
P &=& [1-V_1^{+-}\chi_{11}^{(0)}][1-V_4^{+-}\chi_{44}^{(0)}], \nonumber \\
Q &=& V_1^{+-}\chi_{11}^{(2)}[1-V_4^{+-}\chi_{44}^{(0)}] +
 V_1^{+-} V_4^{+-} |\chi_{14}^{(1)}|^2 + \nonumber \\
   && V_4^{+-}\chi_{44}^{(2)}[1-V_1^{+-}\chi_{11}^{(0)}].
\label{P&Q}
\end{eqnarray}
Setting
 $d(1,4)$ to zero
then yields the approximate energy spectrum of spin excitations,
$\omega_b({\bm q}) = [P({\bm q}) / Q({\bm q})]^{1/2}$,
which is exact in the zero-frequency limit.
Also, applying the RPA solution (\ref{14})
yields the imaginary parts for the dynamical spin susceptibilities of the form
${\rm Im}\,\chi_{11}^{+-}({\bm q},\omega)\cong A_{11}({\bm q})\delta[\omega - \omega_b({\bm q})]$
and
${\rm Im}\,\chi_{44}^{+-}({\bm q},\omega)\cong A_{44}({\bm q})\delta[\omega - \omega_b({\bm q})]$,
with respective spectral weights
\begin{eqnarray}
A_{11} = {\pi\over 2}\sqrt{1-V_4^{+-}\chi_{44}^{(0)}\over{1-V_1^{+-}\chi_{11}^{(0)}}}
{\chi_{11}^{(0)}\over{Q^{1/2}}} , \nonumber \\
A_{44} = {\pi\over 2}\sqrt{1-V_1^{+-}\chi_{11}^{(0)}\over{1-V_4^{+-}\chi_{44}^{(0)}}}
{\chi_{44}^{(0)}\over{Q^{1/2}}}.
\label{A}
\end{eqnarray}
Similar formulae describe the spin dynamics of
 the secondary SDW order parameter $(2,3)$.
Again, we expand the relevant bare spin susceptibilities to lowest non-trivial order in frequency:
\begin{eqnarray}
\chi_{22}^{(0)+-}({\bm q},\omega) &\cong & \chi_{22}^{(0)}({\bm q}) + \omega^2 \chi_{22}^{(2)}({\bm q}),\\
\chi_{23}^{(0)+-}({\bm q},\omega) &\cong & \omega \chi_{23}^{(1)}({\bm q}),\\
\chi_{33}^{(0)+-}({\bm q},\omega) &\cong & \chi_{33}^{(0)}({\bm q}) + \omega^2 \chi_{33}^{(2)}({\bm q}),
\end{eqnarray}
where $\chi_{22}^{(0)}({\bm q}) =\chi_{22}^{(0)+-}({\bm q},0)$,
where $\chi_{33}^{(0)}({\bm q}) =\chi_{33}^{(0)+-}({\bm q},0)$,
where
\begin{eqnarray}
\chi_{22}^{(2)}({\bm q}) = {1\over{\cal N}} 
&\sum_{\bm k}& \biggl(1-{\varepsilon_+\varepsilon_+^{\prime}+\Delta\Delta^{\prime}\over{E E^{\prime}}}\biggr)
{1\over{(E+E^{\prime})^3}} \nonumber \\
&& \cdot {1\over 2}[1-(\cos 2\delta)(\cos 2\delta^{\prime})-(\sin 2\delta)(\sin 2\delta^{\prime})], \\
\chi_{33}^{(2)}({\bm q}) = {1\over{\cal N}} 
&\sum_{\bm k}& \biggl(1+{\varepsilon_+\varepsilon_+^{\prime}-\Delta\Delta^{\prime}\over{E E^{\prime}}}\biggr)
{1\over{(E+E^{\prime})^3}} \nonumber \\
&& \cdot {1\over 2}[1+(\cos 2\delta)(\cos 2\delta^{\prime})-(\sin 2\delta)(\sin 2\delta^{\prime})],
\end{eqnarray}
and where
\begin{equation}
\chi_{23}^{(1)}({\bm q}) = -  {1\over{\cal N}}
\sum_{\bm k}  \biggl({\Delta\over{E}}-{\Delta^{\prime}\over{E^{\prime}}}\biggr){1\over 2}
{\sin(2\delta) - \sin(2\delta^{\prime})\over{(E+E^{\prime})^2}}.
\end{equation}
Again, recall that we have enumerated the indices for the hidden spin and for the
true-SDW moment by $2 = (\pi,0)$ and by $3 = (0,1)$, respectively.
The results for the low-energy spectrum of spin excitations is then identical in form to the
previous ones, (\ref{P&Q}) and (\ref{A}),
but with the replacements 
of the true spin with the hidden spin,
$1\rightarrow 2$,
and with the replacement of the primary hSDW order parameter
with the secondary SDW order parameter,
$4\rightarrow 3$.

\begin{figure}
\includegraphics[scale=0.50, angle=0]{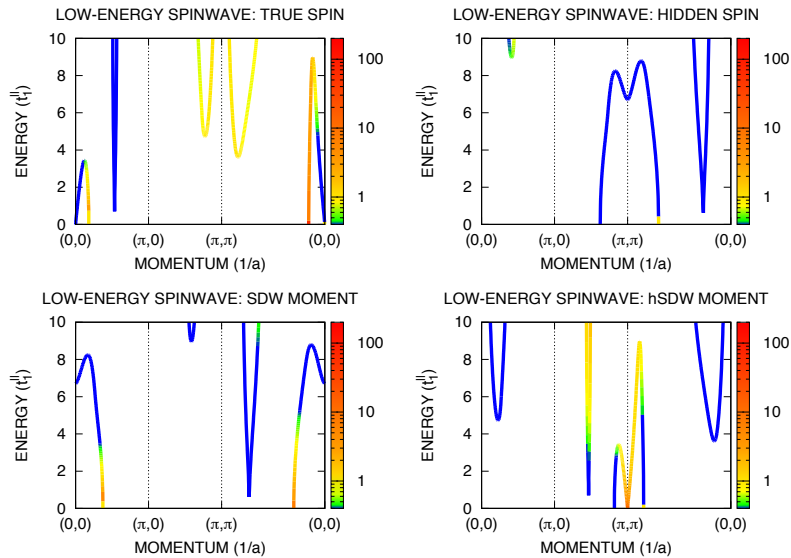}
\caption{Low-energy spectrum of spin excitations predicted by RPA
over a periodic lattice of $1000\times 1000$ iron atoms.
Spectral weight is represented by the color code.
Hopping parameters are listed in the caption to Fig. \ref{FS0},
while Hund and spin-exchange couplings are set to
$J_0 = -100$ meV, $J_1 = 100$ meV, and $J_2 = 50$ meV.
The gap maximum is set to $\Delta_0 = 740$ meV, which implies
$U_0 = 7.37$ eV by the gap equation,
Eq. (\ref{gap_eq}).}
\label{spctrm}
\end{figure}

Figure \ref{spctrm} displays the spectrum of spin excitations for momenta
along a principal axis that is predicted by the low-frequency approximation above.
Hopping matrix elements are set to
 $t_1^{\parallel} = 100$ meV, $t_1^{\perp} = 500$ meV, $t_2^{\parallel} = 0$, and $t_2^{\perp} / i = 100$ meV,
while super-exchange coupling constants are set to
$J_1 = 100$ meV and $J_2 = 50$ meV.
Also,
the  Hund's Rule coupling is set to $J_0 = - 100$ meV,
while the maximum gap is set to $\Delta_0 = 740$ meV.
The gap equation (\ref{gap_eq}) thereby implies a   Hubbard repulsion $U_0 = 7.37$ eV.
Notice that in Fig. \ref{spctrm},
the momenta of the dynamical spin susceptibility within the RPA,
(\ref{14}) and (\ref{23}),
have been shifted by the antiferromagnetic nesting vector,
${\bm Q}_{\rm AF} = (\pi/a, \pi/a)$,
for the true SDW-type and for the hidden SDW-type spin excitations.
They emerge as poles
in frequency of $\chi_{33}^{+-}$ and of $\chi_{44}^{+-}$, respectively.
The latter hSDW-type excitations notably exhibit the expected Goldstone modes
that disperses acoustically from ${\bm Q}_{\rm AF}$.
[See Eq. (\ref{chi_44}) and ref. \cite{jpr_rm_18}.]
By contrast, true SDW-type excitations are predicted by RPA near zero momentum at high energy,
but they have low spectral weight.

Figure \ref{spctrm} also displays moderately strong
excitations near the antiferromagnetic wavevector ${\bm Q}_{\rm AF}$
in the true-spin channel at high energy.  
Below, we will see that they
form a ``floating ring'' of spin excitations around ${\bm Q}_{\rm AF}$.
The lowest-energy ones lie along the diagonal axes of the Brillouin zone.
The latter minima  of this  energy band approach zero energy as the
Hund's Rule coupling, $|J_0|$, increases.
For example,
using the set of parameters that correspond to 
the spin-excitation spectrum displayed by Fig. \ref{spctrm},
while maintaining the gap maximum fixed at $\Delta_0 = 740$  meV,
the lowest-energy of this  band ``touches down'' to zero energy at
a Hund's Rule coupling of $J_{0c} = - 680$ meV, with a Hubbard repulsion of $U_0 = 7.66$ eV.
It is a signal of a quantum phase transition to a different state that obeys Hund's Rule,
such as the conventional SDW state with nesting vector ${\bm Q}_{\rm AF}$.
The spectrum corresponding to true SDW-type excitations remains unchanged, however,
as well as that corresponding to excitations in the hidden-spin channel.
Further, increasing the magnetic frustration, $J_2$, from this point in parameter space
moves back up in energy the ``floating'' ring of magnetic excitations above zero.
This confirms the expectation based on the Heisenberg model that magnetic frustration stabilizes
the hSDW state versus the true SDW state\cite{jpr_10}.

\subsection{General Wavenumbers and Frequency}
\begin{figure}
\includegraphics[scale=0.50, angle=0]{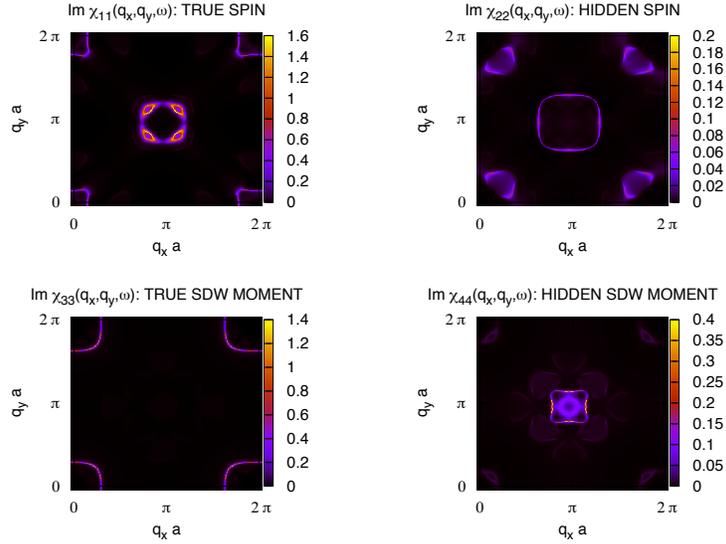}
\caption{Spin excitations at frequency $\omega = 350$ meV
and damping rate $\Gamma = 16$ meV  predicted by RPA
over a periodic lattice of $300\times 300$ iron atoms.
Hopping parameters are listed in the caption to Fig. \ref{FS0},
while Hund and spin-exchange couplings 
and the Hubbard $U_0$ are listed in the caption to Fig. \ref{spctrm}.}
\label{im_chi_3_5}
\end{figure}

\begin{figure}
\includegraphics[scale=0.50, angle=0]{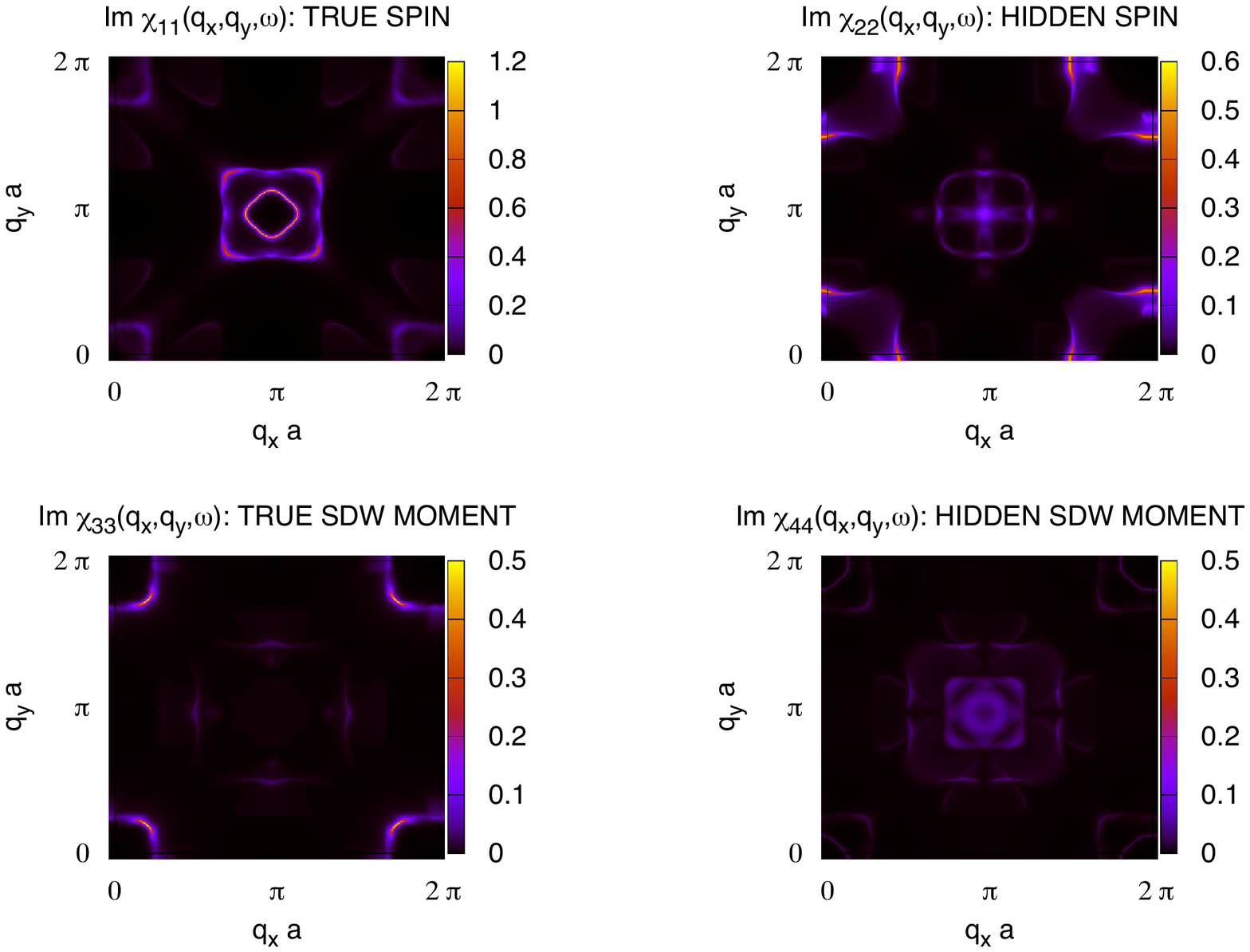}
\caption{Spin excitations at frequency $\omega = 500$ meV
 and damping rate $\Gamma = 16$ meV  predicted by RPA
over a periodic lattice of $300\times 300$ iron atoms.
Hopping parameters,
Hund and spin-exchange couplings,
and the Hubbard $U_0$ are identical to those used in Fig. \ref{spctrm}.}
\label{im_chi_5_0}
\end{figure}

We shall now evaluate the RPA for the dynamical spin susceptibility (\ref{RPA}) numerically
at a fixed frequency $\omega$ and at an artificial damping rate $\Gamma$.
In particular, the explicit expressions for the bare dynamical spin susceptibility
(\ref{chi_0000})-(\ref{chi_p1p1}) are evaluated numerically at complex frequency $\omega + i\Gamma$.
Figure \ref{im_chi_3_5} gives the imaginary part of $\chi^{+-}({\bm q},\omega+i\Gamma)$ at
$\omega = 350$ meV and $\Gamma = 16$ meV.  Hopping parameters and interaction parameters are the same
as those used in Fig. \ref{FS0} and in Fig. \ref{spctrm}.
A smaller periodic square lattice of iron atoms was used, however,
with dimensions $300\times 300$.
And like in Fig. \ref{spctrm},
the momenta of the dynamical spin susceptibility 
have been shifted by the antiferromagnetic nesting vector,
${\bm Q}_{\rm AF} = (\pi/a, \pi/a)$, in the cases of
the true-SDW and of the hidden-SDW channels.
  Notice the moderately strong excitations around the antiferromagnetic wavevector
${\bm Q}_{\rm AF}$ in the true-spin channel.
They emerge near this frequency, and they therefore  coincide with the bottom of the high-energy
bands predicted by the low-frequency approximation above, at  wavenumbers ${\bm q}$ along
a principal axis. (See Fig. \ref{spctrm}.)
Notice also the vestiges of the Goldstone mode centered at ${\bm Q}_{\rm AF}$ in the
hSDW channel.
Figure \ref{im_chi_5_0} shows ${\rm Im}\,\chi^{+-}({\bm q},\omega+i\Gamma)$
 at the same artificial damping rate, $\Gamma = 16$ meV,
but at higher frequency, $\omega = 500$ meV.
The Goldstone mode is hardly visible in the hSDW channel,
but the high-energy spin excitations in the true-spin channel
that circle ${\bm Q}_{\rm AF}$ persist.  
Notice that they now have a ``diamond'' shape.
In summary, the spin-excitation spectrum shows level repulsion at
$\omega\sim 300$ meV, which separates Goldstone modes in the hSDW channel at
low energy from high-energy modes in the true-spin channel.  
Both of these types of spin excitations
are centered at the antiferromagnetic wavevector, ${\bm Q}_{\rm AF}$.

Figure \ref{im_chi_3_5} also shows spin excitations around ${\bm Q}_{\rm AF}$
in the hidden-spin channel and spin-excitations at the center of the Brillouin zone
in the true-SDW channel.  As Fig. \ref{spctrm}
 indicates,
these are related by zone-folding because of the hSDW background,
and they are therefore one and the same.
Figure \ref{im_chi_5_0} displays that such spin excitations no longer exist at higher energy, however.
This  is consistent with the collapsed-dome-shaped band at the center of the folded Brillouin zone
that is
suggested by the low-frequency approximation, Fig. \ref{spctrm},
in  the hidden-spin  and true-SDW channels.

\subsection{Comparison with Heisenberg Model}
The hSDW state studied above was originally discovered in a local-moment model
over a square lattice of iron atoms 
that contain the principal $d+$ and $d-$ orbitals\cite{jpr_17,jpr_19,jpr_ehr_09,jpr_10}.
The model includes Hund's Rule coupling like in $H_U$ (\ref{U})
and Heisenberg exchange coupling like in $H_{\rm sprx}$ (\ref{sprx}).
However, separate intra-orbital versus inter-orbital exchange coupling constants,
$J_1^{\parallel}$ and $J_1^{\perp}$,
exist across nearest neighbors.
Spin-wave theory yields that they are related to the spin stiffness of the hSDW state
by $J_1^{\parallel} - J_1^{\perp} = \rho_s/2s_0^2$. (STRIKE OUT!)
Spin-wave theory also predicts\cite{jpr_rm_18}
 a ``floating ring'' of observable spin excitations around ${\bm Q}_{\rm AF}$.
However, as Hund's Rule coupling $-J_0$ increases, 
the hSDW is eventually destabilized 
by a ``stripe'' SDW that intervenes.
By comparison,
the above RPA calculation does not indicate
that the hSDW state is destabilized by the ``stripe SDW state
 as Hund's Rule is enforced.
This discrepancy 
could be due to the fact that
the local-moment model assumes infinitely strong Hubbard repulsion, $U_0$,
while keeping the Hund's Rule coupling, $-J_0$, finite. (STRIKE OUT!)
Nonetheless, both the present RPA treatment and the previous local-moment model
find that the hSDW state eventually becomes unstable as Hund's Rule is enforced,
as expected.



\section{Discussion and Conclusions} 
Inelastic neutron scattering studies of alkali-atom-intercalated FeSe and 
of organic-molecule-intercalated FeSe find evidence for low-energy spin-excitations
not at,
but around the wavevector $(\pi/a , \pi/a)$ in the unfolded (one-iron) Brillouin zone
\cite{park_11,friemel_12,davies_16,pan_17,ma_17}.
In particular, the lowest-energy spin excitations that  have been observed in
the superconducting state lie just below the gap in energy for quasi-particle excitations,
$2 \Delta_{\rm SC} \cong 28$ meV,
at wavevectors that lie midway between
that corresponding to ``stripe'' SDW order and that corresponding to N\'eel order.
Interestingly, evidence exists  for low-energy spin excitations 
 in the normal state of such electron-doped iron selenides\cite{pan_17}, at wavenumbers near $(\pi/a , \pi/a)$.
In particular,
spin excitations that form a ``diamond'' around this wavevector exist at
energy scales above the gap
 in organic-molecule-intercalated iron-selenide
high-temperature superconductors\cite{pan_17}.

The hSDW state studied here ideally exists at half filling.  It may therefore
provide a good description of the normal state in electron-doped FeSe high-$T_c$ superconductors.
Figures \ref{im_chi_3_5} and \ref{im_chi_5_0} summarize the predictions
 for the nature of high-energy spin excitations
within RPA.  The true-spin channel is most likely the only one that is observable by neutron scattering.
It shows a ``floating ring'' of spin excitations around the antiferromagnetic wavevector
${\bm Q}_{\rm AF}$ that begins at a threshold energy (Fig. \ref{im_chi_3_5}), 
followed by spin excitations at higher energy
that form  a ``diamond'' around the same wavevector (Fig. \ref{im_chi_5_0}).
Inelastic neutron scattering on electron-doped iron selenide indicates that 
relatively high-energy magnetic resonances exist above the quasi-particle energy gap,
in the range $80-130$ meV,
at wavenumbers that roughly form a diamond around the same wavevector\cite{pan_17}.
Note that the threshold energy of the ``floating ring'' shown by Fig. \ref{im_chi_3_5}
 is three times larger.  It can be reduced, however, by increasing the Hund's Rule coupling
towards $|J_{0c}|$,
at which the threshold collapses to zero energy.
The qualitative agreement of theory with experiment suggests that hidden magnetic order
of the type studied here exists in electron-doped iron selenide.

In summary, we have studied the nature of low-energy spin excitations
due to hidden magnetic order in an extended Hubbard model
for a single layer of iron selenide.  
The Hubbard model notably contains only
the two principal $3 d_{xz}$ and $3 d_{yz}$ orbitals of the iron atom.
An RPA was developed
along the lines of the ``spin-bag'' calculation for the
SDW state of the conventional Hubbard model over the square lattice
by Schrieffer, Wen and Zhang\cite{hirsch_85,schrieffer_wen_zhang_89,singh_tesanovic_90,chubokov_frenkel_92}.
It predicts an observable ``diamond'' of spin-excitations around the nesting
vector of the hSDW state, ${\bm Q}_{\rm AF} = (\pi/a, \pi/a)$,
at energies above the band of Goldstone modes,
which are not observable.
Such ``hollowed-out''  spin excitations at ${\bm Q}_{\rm AF}$ have been observed
by inelastic neutron scattering in bulk electron-doped iron-selenide\cite{friemel_12,pan_17}.
The present RPA calculations also predict that they move down in energy as Hund's Rule is enforced,
while that they move up in energy with increasing magnetic frustration.

Absent from the mean-field/RPA study of the hSDW state presented above is 
a description of the superconducting state in electron-doped iron selenide.
Maier and co-workers have proposed that a nodeless $D$-wave paired state
accounts for the spin resonances that lie at energies inside the quasi-particle gap
in electron-doped iron selenide\cite{friemel_12,pan_17,maier_11}.
Mazin argued, however, that a true node appears after zone-folding the one-iron Brillouin zone
because of hybridization between the two inequivalent iron sites in\cite{mazin_11} FeSe.
ARPES finds no evidence for gap nodes\cite{peng_14,lee_14},
on the other hand.
The author has recently found an instability to $S$-wave pairing in the hSDW state upon electron doping,
where the sign of the Cooper pair wavefunction alternates between electron pockets
and faint hole pockets \cite{jpr_20b}.
Such electron/hole pockets lie at the corner of the folded Brillouin zone,
and they are due to a Lifshitz transition of the Fermi surfaces
that is incited by fluctuation-exchange with the Goldstone modes associated with hidden magnetic order
(\ref{chi_44}). [See Fig. \ref{FS1} and ref. \cite{jpr_rm_18}.]
It remains to be seen what type of low-energy spin resonance
 is predicted by such an $S^{+-}$ paired state.

\begin{acknowledgments}
The author would like to thank Jesus P\'erez-Conde for correspondence.
This work was supported in part by the US Air Force
Office of Scientific Research under grant no. FA9550-17-1-0312.
\end{acknowledgments}

\appendix

\section{Orbital Matrix Elements}\label{ppndx_m_e}
The operators that create
the eigenstates (\ref{plane_waves}) of the
electron hopping Hamiltonian, $H_{\rm hop}$, are
\begin{equation}
c_s^{\dagger}(n,{\bm k}) = {\cal N}^{-1/2} \sum_i \sum_{\alpha=0,1}
(-1)^{\alpha n} e^{i(2\alpha-1)\delta(\bm k)} e^{i{\bm k}\cdot{\bm r}_i} c_{i,\alpha,s}^{\dagger},
\label{ck}
\end{equation}
where $\alpha = 0$ and $1$ index the $d-$ and $d+$ orbitals,
and where $n=1$ and $2$ index the anti-bonding and bonding orbitals
$(-i) d_{y(\delta)z}$ and $d_{x(\delta)z}$.
The inverse of the above is then
\begin{equation}
c_{i,\alpha,s}^{\dagger} = {\cal N}^{-1/2} \sum_{\bm k} \sum_{n=1,2}
(-1)^{\alpha n} e^{-i(2\alpha-1)\delta({\bm k})} e^{-i{\bm k}\cdot{\bm r}_i} c_s^{\dagger}(n,{\bm k}).
\label{ci}
\end{equation}
Plugging (\ref{ci}) and its hermitian conjugate into the expression
for the electron spin-density wave operator,
\begin{eqnarray}
{\bm S} (m\pi,{\bm q}) = {1\over 2}\sum_s\sum_{s^{\prime}}
\sum_i\sum_{\alpha} (-1)^{m\alpha} e^{i{\bm q}\cdot{\bm r}_i}
c_{i,\alpha,s}^{\dagger} {\boldsymbol \sigma}_{s,s^{\prime}} c_{i,\alpha,s^{\prime}},
\label{S}
\end{eqnarray}
yields the form
\begin{eqnarray}
{\bm S} (m\pi,{\bm q}) = {1\over 2}\sum_{s}\sum_{s^{\prime}}\sum_{\bm k}\sum_{n,n^{\prime}}
{\cal M}_{n,{\bm k};n^{\prime},{\bm k^{\prime}}} \, c_s^{\dagger}(n^{\prime},{\bm k^{\prime}})
{\boldsymbol \sigma}_{s,s^{\prime}} c_{s^{\prime}} (n,{\bm k}),\nonumber\\
\label{s}
\end{eqnarray}
with matrix element
\begin{equation}
{\cal M}_{n,{\bm k};n^{\prime},{\bm k}^{\prime}} = {1\over 2} \sum_{\alpha=0,1}
e^{i (2 \alpha -1)[\delta({\bm k}) -\delta({\bm k}^{\prime})]} (-1)^{(n^{\prime}-n+m)\alpha}.
\label{A_M}
\end{equation}
Here, $m=0$ or $1$, and ${\bm k}^{\prime} = {\bm k} + {\bm q}$.
The matrix element therefore equals\cite{jpr_rm_18}
\begin{equation}
{\cal M}_{n,{\bm k};n^{\prime},{\bm k}^{\prime}} =
\begin{cases}
\cos[\delta({\bm k})-\delta({\bm k}^{\prime})] & {\rm for}\quad n^{\prime} = n + m\; ({\rm mod}\; 2),\\
-i\, \sin[\delta({\bm k})-\delta({\bm k}^{\prime})] & {\rm for}\quad n^{\prime} = n + m + 1\; ({\rm mod}\; 2).
\end{cases}
\label{A_MM}
\end{equation}
%

\section{Trace Formulas for Products of $2\times 2$ Matrices}\label{ppndx_traces}
Below, we compute the trace of the product of $2\times 2$ matrices
${\rm tr}(\tau_{\mu} \tau_{\gamma} \tau_{\nu} \tau_{\delta})$,
where $\tau_0$ is the identity matrix, and where $\tau_1$, $\tau_2$ and $\tau_3$ are Pauli matrices.
The indices $\mu$ and $\nu$ pertain to the Nambu-Gorkov Greens function:
$G = \sum_{\mu = 0}^{3} G^{(\mu)} \tau_{\mu}$.
We will exploit the product rule
obeyed by Pauli matrices:
\begin{equation}
\tau_i \tau_j = \delta_{i,j} \tau_0 + i\sum_{k=1}^{3} \epsilon_{i,j,k} \tau_k.
\end{equation}
Throughout, greek-letter indices run through $0$, $1$, $2$, and $3$,
while latin-letter indices run through $1$, $2$, and $3$. 

\begin{enumerate}
\item 
${\rm tr}(\tau_{\mu} \tau_{0} \tau_{\nu} \tau_{0}) = 
{\rm tr}(\tau_{\mu} \tau_{\nu}) = 2\, \delta_{\mu,\nu}.$

\item 
${\rm tr}(\tau_{\mu} \tau_{i} \tau_{\nu} \tau_{i}) = 2\, {\rm sgn}_{\mu} (i) \delta_{\mu,\nu}$, \\
where ${\rm sgn}_{\mu} (i) = 1$ if $\mu = 0$ or $i$,
and where ${\rm sgn}_{\mu} (i) = -1$ otherwise.

\item
${\rm tr}(\tau_{\mu} \tau_{0} \tau_{\nu} \tau_{i}) =
 {\rm tr}(\tau_{\mu} \tau_{\nu} \tau_{i}) =
2 (\delta_{\mu,0} \delta_{\nu,i} + \delta_{\mu,i}\delta_{\nu,0}  + i\, \epsilon_{\mu,\nu,i})$, \\
where $\epsilon_{\mu,\nu,i}$ coincides with the Levi-Civita tensor for $\mu,\nu = 1, 2, 3$,
while it vanishes otherwise, for $\mu = 0$, or for $\nu = 0$.

\item
${\rm tr}(\tau_{\mu} \tau_{i} \tau_{\nu} \tau_{0}) =
 {\rm tr}(\tau_{\mu} \tau_{i} \tau_{\nu}) =
2 (\delta_{\mu,0} \delta_{\nu,i} + \delta_{\mu,i}\delta_{\nu,0}  + i\, \epsilon_{\mu,i,\nu})$, \\
where $\epsilon_{\mu,i,\nu}$ coincides with the Levi-Civita tensor for $\mu,\nu = 1, 2, 3$,
while it vanishes otherwise, for $\mu = 0$, or for $\nu = 0$.

\item
${\rm tr}(\tau_{\mu} \tau_{i} \tau_{\nu} \tau_{j}) =
2 (\delta_{\mu,i} \delta_{\nu,j} + \delta_{\mu,j}\delta_{\nu,i}  + 
i\, \delta_{\mu,0}\, \epsilon_{i,\nu,j} + i\, \delta_{\nu,0}\, \epsilon_{\mu,i,j} )$
for $i \neq j$, \\
where $\epsilon_{i,\nu,j}$ and $\epsilon_{\mu,i,j}$ coincide 
with the Levi-Civita tensor for $\mu,\nu = 1, 2, 3$,
while they vanish otherwise, for $\mu = 0$, or for $\nu = 0$.

\end{enumerate}

Importantly, notice that the matrix formed by the trace as a function of
the indices $\gamma$ and $\delta$ is hermitian:
$${\rm tr}(\tau_{\mu} \tau_{\gamma} \tau_{\nu} \tau_{\delta}) =
{\rm tr}(\tau_{\mu} \tau_{\delta} \tau_{\nu} \tau_{\gamma})^* .$$


\clearpage


\begin{thebibliography}{}

\bibitem{keimer_anderson_95} H. F. Fong, B. Keimer, P. W. Anderson, D. Reznik, F. Dogan, and I. A. Aksay,
``Phonon and Magnetic Neutron Scattering at 41 meV in YBa$_2$Cu$_3$O$_7$'',
Phys. Rev. Lett. {\bf 75}, 316 (1995).

\bibitem{inosov_09} D.S. Inosov, J.T. Park, P. Bourges, D.L. Sun, Y. Sidis,
 A. Schneidewind, K. Hradil, D. Haug, C. T. Lin, B. Keimer, and V. Hinkov,
``Normal-State Spin Dynamics and Temperature-Dependent Spin-Resonance Energy
 in Optimally doped BaFe$_{1.85}$Co$_{0.15}$As$_2$''
Nature Physics {\bf 6}, 178 (2010).

\bibitem{korshunov_eremin_08} M.M. Korshunov and I. Eremin,
``Theory of Magnetic Excitations in Iron-Based Layered Superconductors'',
 Phys. Rev. B {\bf 78}, 140509(R) (2008).

\bibitem{maier_scalapino_08} T.A. Maier and D.J. Scalapino,
``Theory of Neutron Scattering as a Probe of the Superconducting Gap in the Iron Pnictides '',
 Phys. Rev. B {\bf 78} 020514(R) (2008).

\bibitem{mazin_08} I.I. Mazin, D.J. Singh, M.D. Johannes, and M.H. Du,
``Unconventional Superconductivity with a Sign Reversal in the Order Parameter of 
LaFeAsO$_{1-x}$F$_x$'',
Phys. Rev. Lett. {\bf 101}, 057003 (2008).

\bibitem{kuroki_08} K. Kuroki, S. Onari, R. Arita, H. Usui, Y. Tanaka, H. Kontani, and H. Aoki,
``Unconventional Pairing Originating from the Disconnected Fermi Surfaces of Superconducting 
LaFeAsO$_{1-x}$F$_x$'',
Phys. Rev. Lett. {\bf 101}, 087004 (2008).

\bibitem{graser_09} S. Graser, T.A. Maier, P.J. Hirschfeld, and D.J. Scalapino,
``Near-Degeneracy of Several Pairing Channels in Multiorbital Models for the Fe Pnictides'',
New J. Phys. {\bf 11}, 025016 (2009).

\bibitem{park_11} J.T. Park, G. Friemel, Yuan Li, J.-H. Kim, V. Tsurkan, J. Deisenhofer,
 H.-A. Krug von Nidda, A. Loidl, A. Ivanov, B. Keimer, and D. S. Inosov,
``Magnetic Resonant Mode in the Low-Energy Spin-Excitation Spectrum of Superconducting 
Rb$_2$Fe$_4$Se$_5$ Single Crystals'',
Phys. Rev. Lett. {\bf 107}, 177005 (2011).

\bibitem{friemel_12} G. Friemel, J.T. Park, T. A. Maier, V. Tsurkan, Yuan Li, J. Deisenhofer, 
H.-A. Krug von Nidda, A. Loidl, A. Ivanov, B. Keimer, and D.S. Inosov,
``Reciprocal-Space Structure and Dispersion of the Magnetic Resonant Mode in the
 Superconducting Phase of Rb$_x$Fe$_{2-y}$Se$_2$ Single Crystals'',
Phys. Rev. B {\bf 85}, 140511(R) (2012).

\bibitem{davies_16}  N.R. Davies, M.C. Rahn, H.C. Walker, 
R.A. Ewings, D.N. Woodruff, S.J. Clarke, and A.T. Boothroyd,
``Spin Resonance in the Superconducting State of Li$_{1-x}$Fe$_x$ODFe$_{1-y}$Se
 Observed by Neutron Spectroscopy'',
 Phys. Rev. B {\bf 94}, 144503 (2016).

\bibitem{pan_17} B. Pan, Y. Shen, D. Hu, Y. Feng, J.T. Park, A.D. Christianson,
 Q. Wang, Y. Hao, H. Wo, Z. Yin, T.A. Maier and J. Zhao,
``Structure of Spin Excitations in Heavily Electron-Doped
Li$_{0.8}$Fe$_{0.2}$ODFeSe Superconductors'',
 Nat. Comm. {\bf 8}, 123 (2017).

\bibitem{ma_17} M. Ma, L. Wang, P. Bourges, Y. Sidis, S. Danilkin, and Y. Li,
``Low-energy Spin Excitations in (Li$_{0.8}$Fe$_{0.2}$)ODFeSe
Superconductor Studied with Inelastic Neutron Scattering'',
 Phys. Rev. B {\bf 95}, 100504(R) (2017).

\bibitem{zhou_13} S. He, J. He, W.-H. Zhang, L. Zhao, D. Liu, X. Liu, D. Mou, Y.-B. Ou,
Q.-Y. Wang, Z. Li, L. Wang, Y. Peng, Y. Liu, C. Chen, L. Yu, G. Liu, X. Dong, J. Xhang,
C. Chen, Z. Xu, X. Chen, X. Ma, Q. Xue, and X.J. Zhou,
``Phase Diagram and Electronic Indication of High-Temperature Superconductivity
 at 65 K in Single-Layer FeSe Films'',
Nat. Mater. {\bf 12}, 605 (2013).

\bibitem{peng_14} R. Peng, X.P. Shen, X. Xie, H.C. Xu, S.Y. Tan, M. Xia,
T. Zhang, H.Y. Cao, X.G. Gong, J.P. Hu, B.P. Xie, D. L. Feng,
``Measurement of an Enhanced Superconducting Phase and a Pronounced Anisotropy of 
the Energy Gap of a Strained FeSe Single Layer in FeSe/Nb: SrTiO$_3$/KTaO$_3$
 Heterostructures Using Photoemission Spectroscopy'',
Phys. Rev. Lett. {\bf 112}, 107001 (2014).

\bibitem{lee_14} J.J. Lee, F.T. Schmitt, R.G. Moore, S. Johnston, Y.-T. Cui, W. Li,
M. Yi, Z.K. Liu, M. Hashimoto, Y. Zhang, D.H. Lu,
T.P. Devereaux, D.-H. Lee and Z.-X. Shen,
``Interfacial Mode Coupling as the Origin of the Enhancement of $T_c$ in FeSe Films on SrTiO$_3$'',
 Nature {\bf 515}, 245 (2014).

\bibitem{zhao_16} L. Zhao, A. Liang, D. Yuan, Y. Hu, D. Liu, J. Huang,
S. He, B. Shen, Y. Xu, X. Liu, L. Yu, G. Liu, H. Zhou, Y. Huang, X. Dong,
F. Zhou, Z. Zhao, C. Chen, Z. Xu  and,X.J. Zhou,
``Common Electronic Origin of Superconductivity in (Li,Fe)OHFeSe
Bulk Superconductor and Single-Layer FeSe/SrTiO$_3$ Films'',
Nat. Comm. {\bf 7}, 10608 (2016).

\bibitem{jpr_17} J.P. Rodriguez,
``Isotropic Cooper Pairs with Emergent Sign Changes in a Single-Layer Iron Superconductor'',
 Phys. Rev. B {\bf 95}, 134511 (2017).

\bibitem{jpr_19} J.P. Rodriguez,
``Particle-Hole Transformation in Strongly-Doped Iron-Based Superconductors'',
Symmetry {\bf 11}, 396 (2019).

\bibitem{jpr_rm_18} J.P. Rodriguez and R. Melendrez,
``Fermi Surface Pockets in Electron-Doped Iron Superconductor by Lifshitz Transition'',
 J. Phys. Commun. {\bf 2}, 105011 (2018);
``Corrigendum: Fermi Surface Pockets in Electron-Doped Iron Superconductor by Lifshitz Transition'',
J. Phys. Commun. {\bf 3}, 019501 (2019).

\bibitem{jpr_ehr_09} J.P. Rodriguez and E.H. Rezayi,
``Low Ordered Magnetic Moment by Off-Diagonal Frustration in
 Undoped Parent Compounds to Iron-Based High-$T_c$ Superconductors'',
 Phys. Rev. Lett. {\bf 103}, 097204 (2009).

\bibitem{jpr_10} J.P. Rodriguez,
``Magnetic Excitations in Ferropnictide Materials Controlled by a Quantum Critical Point
into Hidden Order'',
 Phys. Rev. B {\bf 82}, 014505 (2010).

\bibitem{jpr_20b} J.P. Rodriguez,
``Superconductivity by Hidden Spin Fluctuations in Electron-Doped Iron Selenide'',
 arXiv:2001.07908 .

\bibitem{nambu_60} Y. Nambu,
``Quasi-Particles and Gauge Invariance in the Theory of Superconductivity'',
 Phys. Rev. {\bf 117}, 648 (1960).

\bibitem{gorkov_58} L.P. Gorkov, Zh. Eksperim. i Teor. Fiz. {\bf 34}, 735 (1958); 
``About the Energy Spectrum of Superconductors'',
Sov. Phys. JETP {\bf 7}, 505 (1958).

\bibitem{schrieffer_64} J.R. Schrieffer, {\it Theory of Superconductivity}
(Benjamin, New York, 1964).

\bibitem{hirsch_85} J.E. Hirsch,
``Two-Dimensional Hubbard Model: Numerical Simulation Study'',
 Phys. Rev. {\bf B} 31, 4403 (1985).

\bibitem{schrieffer_wen_zhang_89} J.R. Schrieffer, X.G. Wen, and S.C. Zhang,
``Dynamic Spin Fluctuations and the Bag Mechanism of High-$T_c$ Superconductivity'',
Phys. Rev. B {\bf 39}, 11663 (1989).

\bibitem{singh_tesanovic_90} A. Singh and Z. Tesanovic,
``Collective Excitations in a Doped Antiferromagnet'',
Phys. Rev. B {\bf 41}, 614 (1990).

\bibitem{chubokov_frenkel_92} A.V. Chubukov and D.M. Frenkel,
``Renormalized Perturbation Theory of Magnetic Instabilities
in the Two-Dimensional Hubbard Model at Small Doping'',
Phys. Rev. B {\bf 46}, 11884 (1992).

\bibitem{raghu_08} S. Raghu, Xiao-Liang Qi, Chao-Xing Liu, D.J. Scalapino, Shou-Cheng Zhang,
``Minimal Two-Band Model of the Superconducting Iron Oxypnictides'',
Phys. Rev. B {\bf 77}, 220503(R) (2008).

\bibitem{Lee_Wen_08} P.A. Lee and X.-G. Wen,
``Spin-Triplet P-Wave Pairing in a Three-Orbital Model for Iron Pnictide Superconductors'',
 Phys. Rev. B {\bf 78}, 144517 (2008).

\bibitem{jpr_mana_pds_14} J.P. Rodriguez, M.A.N. Araujo and P.D. Sacramento,
``Emergent Nesting of the Fermi Surface from
 Local-Moment Description of Iron-Pnictide High-$T_c$ Superconductors'',
Eur. Phys. J. B {\bf 87}, 163 (2014).

\bibitem{anderson_50} P.W. Anderson,
``Antiferromagnetism. Theory of Superexchange Interaction'',
Phys. Rev. {\bf 79}, 350 (1950).

\bibitem{Si&A} Q. Si and E. Abrahams, 
``Strong Correlations and Magnetic Frustration in the High-$T_c$ Iron Pnictides'',
Phys. Rev. Lett. {\bf 101}, 076401 (2008).

\bibitem{2orb_Hbbrd} M. Daghofer, A. Moreo, J.A. Riera, E. Arrigoni, D.J. Scalapino, and E. Dagotto,
``Model for the Magnetic Order and Pairing Channels in Fe Pnictide Superconductors'',
Phys. Rev. Lett. {\bf 101}, 237004 (2008);
A. Moreo, M. Daghofer, J.A. Riera, and E. Dagotto,
``Properties of a Two-Orbital Model for Oxypnictide Superconductors: Magnetic Order, 
B$_{2 g}$ Spin-Singlet Pairing Channel, and its Nodal Structure'',
Phys. Rev. B {\bf 79}, 134502 (2009).

\bibitem{xu_muller_sachdev_08}
C. Xu, M. M\"uller, S. Sachdev,
``Ising and Spin Orders in the Iron-Based Superconductors'',
Phys. Rev. B {\bf 78}, 020501(R) (2008).

\bibitem{yoshizawa_simayi_12}
M. Yoshizawa and S. Simayi,
``Anomalous Elastic Behavior and its Correlation with Superconductivity
in Iron-Based Superconductor Ba(Fe$_{1-x}$Co$_x$)$_2$As$_2$'',
Mod. Phys. Lett. B {\bf 26}, 1230011 (2012).

\bibitem{BMS_12} E. Berg, M.A. Metlitski, and S. Sachdev,
``Sign-Problem-Free Quantum Monte Carlo of the Onset of Antiferromagnetism in Metals'',
 Science {\bf 338}, 1606 (2012).

\bibitem{halperin_hohenberg_69} B. I. Halperin and P. C. Hohenberg, 
``Hydrodynamic Theory of Spin Waves'',
Phys. Rev. {\bf 188}, 898 (1969).

\bibitem{forster_75} D. Forster, {\it Hydrodynamic Fluctuations, Broken Symmetry,
and Correlation Functions} (Benjamin/Cummings, Reading, MA, 1975).

\bibitem{maier_11} T.A. Maier, S. Graser, P.J. Hirschfeld, D.J. Scalapino,
``D-Wave Pairing from Spin Fluctuations in the K$_x$Fe$_{2-y}$Se$_2$ Superconductors'', 
 Phys. Rev. B {\bf 83}, 100515(R) (2011).

\bibitem{mazin_11} I.I. Mazin,
``Symmetry Analysis of Possible Superconducting States in K$_x$Fe$_y$Se$_2$ Superconductors'',
 Phys. Rev. B {\bf 84}, 024529 (2011).

\end{thebibliography}
\end{document}